# Ultrafast electron diffraction imaging of gas-phase molecules

K. Amini[1,*], J. Biegert[1,2,+]

[1]ICFO - Institut de Ciencies Fotoniques, The Barcelona Institute of Science and Technology, 08860 Castelldefels (Barcelona), Spain.
[2]ICREA, Pg. Lluís Companys 23, 08010 Barcelona, Spain.
*kasra.amini@icfo.eu
+jens.biegert@icfo.eu

**Abstract**

Knowledge of molecular structure is paramount in understanding, and ultimately influencing, chemical reactivity. For nearly a century, diffractive imaging has been used to identify the structures of many biologically-relevant gas-phase molecules with atomic (*i.e.* Ångstrom, Å; 1Å = $10^{-10}$ m) spatial resolution. Unravelling the mechanisms of chemical reactions requires the capability to record multiple well-resolved snapshots of the molecular structure as it is evolving on the nuclear (*i.e.* femtosecond, fs; 1 fs = $10^{-15}$ s) timescale. We present the latest, state-of-the-art ultrafast electron diffraction methods used to retrieve the molecular structure of gas-phase molecules with Ångstrom and femtosecond spatio-temporal resolution. We first provide a historical and theoretical background to elastic electron scattering in its application to structural retrieval, followed by details of field-free and field-dressed ultrafast electron diffraction techniques. We discuss the application of these ultrafast methods to time-resolving chemical reactions in real-time, before providing a future outlook of the field and the challenges that exist today and in the future.

# I – Introduction

Chemical reactivity is governed by the location and interaction of electrons and nuclei within atoms and molecules. Pinpointing the location of atoms that typically have an atomic radius of ~0.5 Ångstrom (Å; 1 Å = $10^{-10}$ m) is crucial in determining bond lengths in molecules which typically span 1-3 Å [Weast1984]. For example, the H-H, C-H, and C-C bond lengths in molecular hydrogen ($H_2$) and ethane ($C_2H_6$) are 0.74 Å [Huber1979], 1.09 Å [Herzberg1966], and 1.54 Å [Herzberg1966]. The structure of gas-phase molecules can be determined with atomic (*i.e.* Ångstrom) spatial resolution using a variety of techniques, such as: microwave rotational spectroscopy [Gordy1970, Pratt1998, Schermann2007, Katrizky2010, Laane2011, Patterson2013, Lindon2016, Gordon2017], infrared vibrational spectroscopy [Pratt1998, Larkin2017], Raman spectroscopy [Schrader1995], Coulomb explosion imaging (CEI) [Vager1976, Kanter1979, Vager1986, Naaman1989, Vager1993, Stapelfeldt1995, Stapelfeldt1998, Legare2005, Christensen2014, Christensen2015, Slater2015, Pickering2016, Xu2016, Frasinski2016, Amini2017a, Amini2017b, Burt2017, Ablikim2017, Allum2018, Amini2018, Brausse2018, Burt2018, Karamatskos2019b], and diffraction imaging [Zuo1996, Lein2002, Corkum2007, Meckel2008, Lin2010, Okunishi2011, Xu2012, Blaga2012, Xu2014, Pullen2015, Yu2015, Wolter2016, Ito2016, Pullen2016, Ito2017, Amini2019a, Liu2019, Ueda2019, Amini2019b, Fuest2019, Karamatskos2019a, Hensley2012, Yang2016a, Yang2016b, Centurion2016, Yang2018a, Wolf2019, Shen2019, Bressler2004, Chergui2009, Filsinger2011, Kuepper2014, Sternr2014, Minitti2015, Minemoto2016].

Microwave spectroscopy is the gold-standard of structural retrieval methods used in physical chemistry for determining molecular structures with high precision and resolution. Irradiating a gas-phase sample with microwave radiation leads to transitions between the quantum rotational levels of the polar molecule. From the resulting rotational spectra, molecular structure is retrieved through comparison to corresponding calculated rotational distributions. In vibrational spectroscopy, a gas-phase molecular sample is illuminated with infrared radiation that resonantly excites the vibrational degrees of freedom of the polar molecule possessing a dipole moment. The transmitted light is measured to obtain an absorption vibrational spectrum. The $3N-5$ ($3N-6$) resonantly excited normal modes that appear in the resulting vibrational spectra can be used to identify the geometric structure of the linear (non-linear) molecule. The measured vibrational spectra can also contain contributions from anharmonicities (*e.g.* overtones, combination bands and Fermi resonances) as well as rovibrational coupling that can further split the measured vibrational bands [Wilson1980]. Consequently, the interpretation of vibrational spectra becomes more challenging with increased molecular complexity, and thus typically requiring the use of theoretical calculations to retrieve structural information from the best fit to measured vibrational spectra. Raman spectroscopy enables the structural retrieval of non-polar molecules through the inelastic scattering of incoming photons that are dependent on the change in the molecule's polarizability. The measured inelastically scattered light provides a molecular fingerprint of the measured molecular structure [Schrader1995].

In CEI, several electrons are ionized from the molecule, leading to Coulomb repulsion and subsequent Coulomb explosion between various charged sites of the molecule to generate several positively charged ionic fragments. The recoil velocities of the ionic fragments can be indicative of the starting position of each atom in the molecule at the moment of initial ionization. CEI coupled with covariance imaging [Frasinski2016] has demonstrated the ability to retrieve structural information of polyatomic molecules [Christensen2014, Slater2015, Frasinski2016, Amini2017a, Amini2017b, Burt2018, Karamatskos2019b] so long as the molecule does not significantly change in structure during the Coulomb explosion process, abiding by the "axial recoil approximation" [Christensen2016]. Any perturbations in structure during the Coulomb explosion process can possibly make any non-axial recoil velocities difficult to interpret.

Diffraction imaging can directly retrieve the structure of cells and molecules with Ångström spatial resolution using methods, such as: conventional electron diffraction (CED) [Mark1930, Hargittai1988, Centurion2016], scanning tunnelling microscopy (STM) [Chen2008, Binnig1982], and atomic force microscopy (AFM) [Binnig1986, Gross2009, Zhang2013]. In this chapter, the focus will be on gas-phase diffraction-based imaging techniques, particularly that of electron diffraction. In the following, a brief history of diffraction imaging is given. Diffraction imaging was born from Einstein's 1905 revolutionary concept of light being treated as both a type of electromagnetic wave, possessing a wavelength and frequency, and as behaving like a particle, localized in packets of discrete energy, called photons. Using light's wave-particle duality, Debye's 1915 theoretical study demonstrated that the interference effects from the X-ray diffraction of randomly-oriented electrons and molecules survive and strongly influence the detected averaged intensity distribution as a function of the scattering angle [Debye1915]. Moreover, the experimentally observed Compton scattering [Compton1922a, Compton1922b] in 1922 could only be explained by light's wave-particle duality, confirming Einstein's 1905 hypothesis. Compton scattering occurs when an incoming photon scatters against a charged particle (*e.g.* an electron) to generate a scattered photon with a longer wavelength than it initially had following the transfer of energy from the scattered photon to the recoiled charged particle [Compton1923, Debye1923].

De Broglie' theory of wave-particle duality in 1924 established that in fact any discrete form of matter (*e.g.* electrons) can possess both particle-like behaviour and wave-like properties by establishing a relationship between the velocity, $v$, mass, $m$, and de Broglie wavelength, $\lambda_B$, of the particle, as given by [DeBroglie1924, DeBroglie1925]

$$\lambda_B = \frac{h}{mv} = \frac{h}{\sqrt{2meU}}, \tag{1.1}$$

where $h$ is Planck's constant, $e$ is the elementary charge, and $U$ is the accelerating potential. The wave nature of electrons was experimentally confirmed by the first electron diffraction measurements on crystals in 1927 [Davisson1927, Thomson1927], followed by the first gas-phase electron diffraction measurements in 1930 that demonstrated the successful molecular structure retrieval of carbon tetrachloride, $CCl_4$ [Mark1930]. In these early electron diffraction measurements, the interatomic distances were initially determined by measuring the positions of the maxima and minima on the interference pattern [Wierl1930], before the more direct method of Fourier-transforming the interference signal was adopted soon afterwards to obtain a radial distribution function that is related to the probability distribution of interatomic distances [Pauling1935]. In this radial distribution function, the centre of the peaks represents the average internuclear distance value, whilst the full-width at half maximum (FWHM) corresponds to its vibrational amplitude [James1932]. Gas-phase electron diffraction has successfully identified the static structures of neutral molecules [Allen1950, Sciaini2011, Centurion2016], molecular ions [Turman1968], stable [Bohn1967] and unstable [Schäfer1968] free radicals, and conformers in conformational mixtures [Hassel1943, Bastiansen1946, Mirono1985].

Tracking chemical reactions requires imaging techniques that can not only retrieve the molecular structure with atomic (*i.e.* Ångstrom) spatial resolution but to also monitor changes in its geometric structure on the nuclear (*i.e.* femtosecond) timescale. Fig. 1.1 illustrates the typical timescales and length scales that various different chemical, biological and physical processes proceed with. Time-resolved snapshots of molecular structure can be measured by using the pump-probe technique [Norrish1949, Porter1950, Porter1972, Dantus1987, Dantus1988, Rosker1988, Zewail1988]. Here, an optical (*i.e.* "pump") pulse photoinitiates a chemical reaction of interest, providing the exact start time of the chemical reaction, $t_0$. Then, a second, separate (*i.e.* "probe") pulse captures multiple snapshots of the evolving transient molecular structure at varied times relative to $t_0$, called the pump-probe delay, $\Delta t$. Consequently, a "molecular movie" of the photoinduced chemical reaction can be generated from the accumulated time-resolved snapshots. Recent advances in femtosecond laser and undulator technologies led to the development of femtosecond pulsed sources of X-rays and electrons, and ultimately to the various different femtosecond-resolved diffraction imaging techniques [Bressler2004, Chergui2009, Sciaini2011, Spence2012, Yong2019]. Ultrafast X-ray diffraction (UXD) has become viable through the development of ultrabright, femtosecond optical pulses with high photon energies in the X-ray regime following the commissioning of X-ray free-electron lasers (XFELs) [Chapman2006, Chapman2009, Chapman2011, Spence2012, Minitti2015, Kirrander2017, Sobolev2019, Yong2019]. Such ultrabright, high fluence X-ray sources are required since the photon scattering cross-section is small, making UXD a "photon hungry" method which can damage the sample if the UXD pulse duration is not short enough [Sobolev2019]. Currently, ultrabright UXD measurements are only possible at large-scale, costly, international light source facilities, such as the Linac Coherent Light Source (LCLS) and the European XFEL, access to which is limited to pre-selected beamtimes [Minitti2015, Young2018]. In fact, electron scattering is favoured over photon scattering due to the 5-6 orders-of-magnitude higher scattering cross-section, and the small de Broglie wavelength of electrons that provide picometre spatially-resolved interrogations of molecule structure [Chergui2009]. This is because X-rays are scattered by the electron density distribution surrounding the atoms, whilst electrons are scattered by the electric field of the positively charged nuclei to help pin-point the nuclear core of each atom. Moreover, the much stronger interaction of the electron beam with the target leads to the larger absorption of electrons (beta radiation) than X-rays (gamma radiation). As a result, a wide range of molecular imaging studies are based on electron scattering methods, including: non-relativistic and relativistic ultrafast electron diffraction (UED) [Ihee2001, Siwick2002, Wang2003, Siwick2003, Srinivasan2003, Siwick2004, King2005, Hastings2006, Wang2006, Zewail2006, Sciaini2011, Hensley2012, Yang2016a, Yang2016b, Carbajo2016, Centurion2016, Yang2018a, Wolf2019, Shen2019, Ischenko2019], ultrafast electron microscopy (UEM) [Lobastov2005, King2005, Zewail2006, Hassan2017, Hassan2018], laser-assisted electron diffraction (LAED) [Morimoto2014], and laser-induced electron diffraction (LIED) [Zuo1996, Lein2002, Corkum2007, Meckel2008, Lin2010, Okunishi2011, Xu2012, Blaga2012, Xu2014, Pullen2015, Yu2015, Wolter2016, Ito2016, Pullen2016, Ito2017, Amini2019a, Liu2019, Ueda2019, Amini2019b, Fuest2019, Karamatskos2019a].

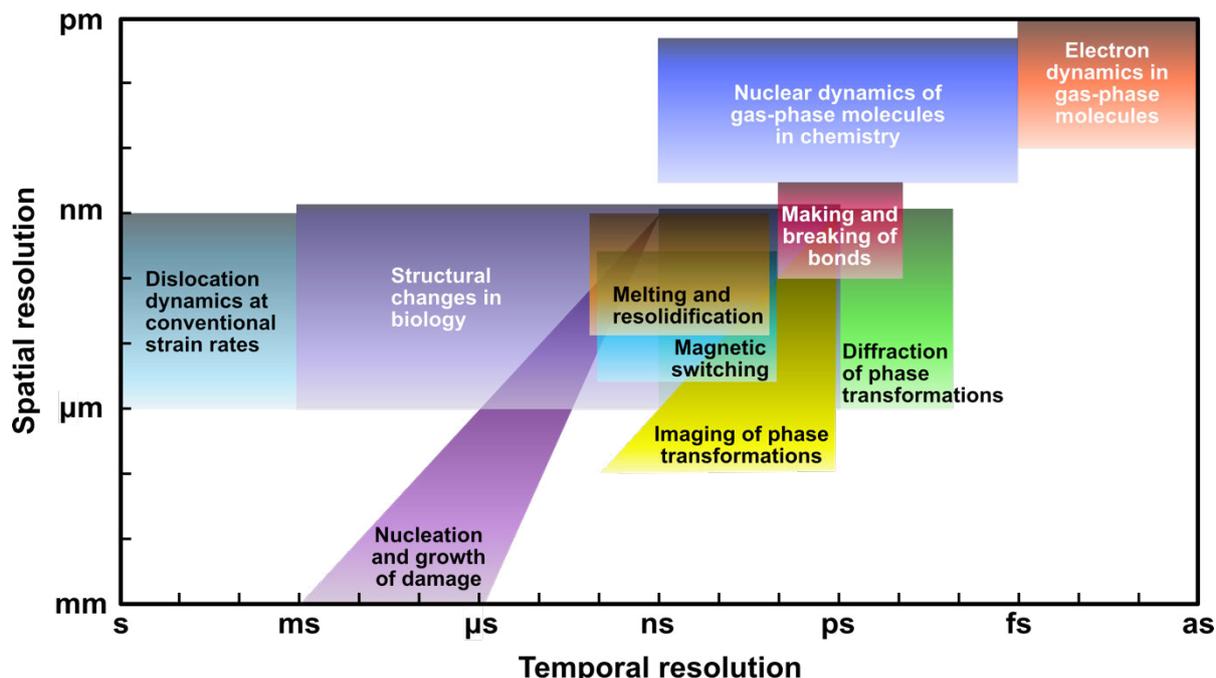

**Fig. 1.1** Various different physical phenomena in chemistry, biology and physics together with their corresponding spatial and temporal resolutions. Figure adapted from [King2005].

This chapter is organised as follows: details on elastic electron scattering is given in Section II, followed by details of ultrafast field-free and field-dressed electron diffraction imaging techniques in Sections III and IV, respectively, before concluding remarks on the future outlook of the electron diffraction field are given in Section V.

**II – Elastic electron scattering**

The geometric structure of gas-phase molecules can be directly retrieved through elastic electron scattering, as illustrated in Fig. 2.1. Here, an incident electron beam travelling along the $z$ axis with a momentum/wave vector $\vec{k}_0$ that impinges onto an exemplary gas-phase diatomic molecule (grey structure at origin of Fig. 2.1). Due to the wave nature of the electron, the incoming electron beam can be described by a planar wave given by,

$$\psi_0 = A e^{ik_0 z}, \qquad (2.1)$$

which has a wavelength of $\lambda = 2\pi/k_0$.

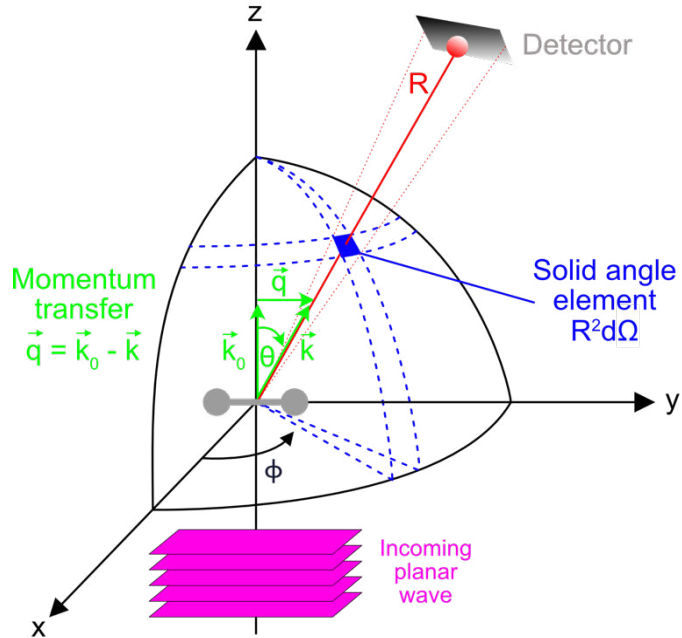

**Fig. 2.1** Illustration of elastic electron scattering in polar coordinates. An incoming planar wave (pink rectangular planes) with initial momentum $\vec{k}_0$ scatters against a generic diatomic molecule (grey structure at origin). After interaction, the electron is scattered with a momentum $\vec{k}$ in the scattering and azimuthal direction given by the angles θ and ϕ, respectively. This leads to a momentum transfer of $\vec{q} = \vec{k}_0 - \vec{k}$ (see green) that is dependent on the molecular structure. The scattered electron is detected on a detector (grey rectangle) that is a distance of $R$ away from the point of interaction, with the solid angle element (blue filled square) given by $R^2 d\Omega$.

As the beam of electrons travels through the electrostatic potential of the molecule, the electron beam is deflected from its initial direction with a scattering angle, $\theta$, that is dependent on the target molecule's geometric structure, and collected within a solid angle Ω at a detector that is a distance of $R$ away from the point of interaction. Here, the elastic scattering of an electron against an atom is treated as a hard collision, and the probability of the collision occurring is given by the scattering cross-section, $\sigma$. In electron diffraction measurements, this cross-section is typically investigated as a function of the electron's angle and energy, leading to a doubly-differential total elastic cross-section, $\sigma_{2D}$. Specifically, the number of electrons scattered into a specific solid angle element is experimentally determined and is referred to as the doubly-differential differential cross-section (DCS), $d^2\sigma/d\Omega^2$. The total cross-section for elastic scattering, $\sigma_{2D}$, can therefore be obtained from the measured DCS by integrating the DCS over all solid angles.

To interrogate the molecular structure without excitation, we consider the electron beam scattering off the target without any energy loss, called elastic electron scattering. The elastically scattered electron beam in the polar coordinate system can be described by a spherical wave given by,

$$\psi' = A \frac{1}{R} f(\theta) e^{ik_0 R}, \qquad (2.2)$$

where $R$ is the distance from the point of scattering to the observation point where the scattered electron was detected, and $f(\theta)$ is the electron scattering amplitude. In electron diffraction experiments, the scattering cross-section, $(I/I_0)R^2$, is in fact measured which is directly related to the scattering amplitude through the scattered electron intensity, $I$, given by

$$I = \frac{I_0}{R^2} |f(\theta)|^2, \qquad (2.3)$$

where $I_0$ is the incident electron intensity. Calculating the electron scattering amplitude will provide a handle on the scattering cross-section, which is embedded in the measured molecular interference signal of electron scattering measurements.

Taking the whole molecule as a one-centre scatterer, the scattered wave is given by

$$\psi'(\vec{R}) = KA \frac{e^{ik_0 R}}{R} \frac{1}{q^2} \left[ \underbrace{\sum_i Z_i e^{i(\vec{q}\vec{r}_i)}}_{\text{nuclei}} - \underbrace{\int_V \frac{1}{e} \rho(\vec{r}) e^{i(\vec{q}\vec{r})} dr}_{\text{electrons}} \right], \qquad (2.4)$$

where $K = 8\pi^2 m e^2/h^2$, the charges and vector coordinates of the nuclei in the molecule are given by $Z_i e$ and $\vec{r}_i$, respectively, and $\rho(\vec{r})$ is the electron density at any point $\vec{r}$ in the molecule. Here, the momentum transfer,

$$q = |\vec{q}| = |\vec{k}_0 - \vec{k}| = 2k_0 \sin\left(\frac{\theta}{2}\right), \qquad (2.5)$$

is the momentum transferred to the scattered electron wave after the incoming electron wave interacted with the target molecule, where $|\vec{k}| = |\vec{k}_0|$ for elastic scattering (see Fig. 2.1 in green). The momentum transfer provides the degree to which the nuclei of a system is "penetrated" by the incoming electron wave. The two terms in the square bracket in Eqn (2.4) correspond to electron scattering by (i) the nuclei (left part) and (ii) the electron density distribution (right part).

Extending electron scattering to a molecule can be achieved by considering elastic scattering occurring locally on each individual atom in the molecule, and so Eqn (2.4) can be modified to determine the atom-dependent scattering amplitudes in a molecular system *via* the independent atom model (IAM) to give the scattered wave

$$\psi'(\vec{R}) = KA \frac{e^{ik_0 R}}{R} \frac{1}{q^2} \sum_i f_i(\theta) e^{i(\vec{q}\vec{r})}, \tag{2.6}$$

where the electron scattering amplitude on atom $i$ is given by

$$f_i(\theta) = \frac{Z_i - F_i}{q^2}, \tag{2.7}$$

and $F_i$ is the atomic scattering factor due to scattering by the electron density distribution at atom $i$ given by

$$F_i = \int_V \rho(\vec{r}_i') e^{i(\vec{q}\vec{r})} d\vec{r}_i'. \tag{2.8}$$

In fact, the scattering amplitude given by Eqn (2.6) encompasses the scattering of electrons by the molecule's nuclei ($Z_i$; Eqn (2.7)) and electrons ($F_i$; Eqn (2.8)). Since photons are only scattered by the molecule's electron distribution, Eqn (2.8) is only needed to characterize X-ray scattering. It should be noted that although the use of spherical waves provides a meaningful physical picture that explains the elastic electron scattering process, more accurate scattering amplitudes based on partial waves are instead typically used in retrieval algorithms. Specifically, partial waves are used to account for the various orbital angular momenta that the scattered wave experiences when interacting with the target molecule, providing a more accurate and quantitative account of the scattering process.

To obtain meaningful structural information of the target molecule, it is useful to consider Eqn (2.6) as a function of the internuclear distance between two atoms $i$ and $j$,

$$\vec{r}_{ij} = \vec{r}_i - \vec{r}_j, \tag{2.9}$$

which act as two-centre scatterers in the molecule. For a static molecule at a well-defined spatial orientation, this leads to the electron scattering intensity to be given as

$$I(q) = \frac{K^2 I_0}{R^2} \sum_{i=1}^{N} \sum_{j=1}^{N} f_i(q) f_i^*(q) e^{i(\vec{q}\vec{r}_{ij})}. \tag{2.10}$$

Typically, in an electron scattering experiment, the molecules are randomly oriented, and in good approximation, each molecular orientation possesses equal likelihood of existing, leading to a detected scattering signal that is averaged over all orientations. To account for randomly oriented molecules, the electron scattering intensity is then modified to

$$I(q) = \frac{K^2 I_0}{R^2} \sum_{i=1}^{N} \sum_{j=1}^{N} f_i(q) f_i^*(q) \int_0^\pi e^{i(\vec{q}\vec{r}_{ij}\cos(\alpha))} \frac{1}{2}\sin(\alpha) d\alpha \tag{2.11}$$

$$I(q) = \frac{K^2 I_0}{R^2} \sum_{i=1}^{N} \sum_{j=1}^{N} f_i(q) f_i^*(q) \left(\frac{\sin(qr_{ij})}{qr_{ij}}\right) \tag{2.12}$$

in order to take into account the particular molecular orientation that appears between the intervals of $\alpha$ and $\alpha + d\alpha$ with a probability of

$$P_\alpha d\alpha = \frac{1}{2}\sin(\alpha) d\alpha, \tag{2.13}$$

where $\alpha$ is the angle between the $\vec{q}$ and $\vec{r}_{ij}$. Using the atomic electron scattering amplitude from the partial waves method,

$$f_i(q) = |f_i(q)|e^{[i\eta_i(q)]}, \tag{2.14}$$

the total electron scattering intensity, $I_T(q)$, averaged over all molecular orientations can be written as a function of the absolute value of the scattering amplitude, $|f_i(q)|$, and its phase for the atom $i$, given by

$$I_T(q) = \left(\frac{K^2 I_0}{R^2}\sum_{i=1}^{N}|f_i(q)^2|\right) + \left(\frac{K^2 I_0}{R^2}\sum_{\substack{i=1\\i\neq j}}^{N}\sum_{\substack{j=1\\i\neq j}}^{N}|f_i(q)||f_j(q)|\cos[\eta_i(q)-\eta_j(q)]\frac{\sin(qr_{ij})}{qr_{ij}}\right) \tag{2.15}$$

$$I_T(q) = I_A(q) + I_M(q), \tag{2.16}$$

which in fact has two contributing factors: an incoherent atomic part, $I_A(q)$, and a coherent molecular part, $I_M(q)$. The physical significance of these two contributions are explained using a diatomic molecule as an exemplary case. The molecular interference signal, $I_M$, arises from the incoming electron beam interacting with the two scattering centres of the diatomic molecule to generate spherical waves (similar to Young's double slit) that can constructively or destructively interfere. Consequently, the $I_M$ is sensitive to the internuclear distance between two scattering centres (analogous to the distance between two slits). Whilst the incoherent atomic part, $I_A$, arises from the scattering of electrons on only a single atom, leading to a background signal that is the incoherent sum of scattering on one-atom centres that contains no structural information. Since the background atomic scattering signal, $I_A$, dominates the measured total scattering signal and to compensate for the significant decrease in scattering intensity with increasing momentum transfer, the molecular interference signal is contrasted against $I_A$ to highlight the interference signal of interest, $I_M$, through the molecular contrast factor (MCF) given by

$$MCF = \frac{I_T - I_A}{I_A} = \frac{I_M}{I_A} \tag{2.17}$$

$$MCF = \frac{1}{\sum_{i=1}^{N}|f_i(q)^2|}\sum_{\substack{i=1\\i\neq j}}^{N}\sum_{\substack{j=1\\i\neq j}}^{N}|f_i(q)||f_j(q)|\cos[\eta_i(q)-\eta_j(q)]\frac{\sin(qr_{ij})}{qr_{ij}}. \tag{2.18}$$

As is evident from Eqn (2.18), each internuclear distance in the molecule will have a sinusoidal response. Fig. 2.2A shows a sinusoidal calculated MCF distribution for the triatomic molecule carbon dioxide ($CO_2$). Here, the $CO_2$ MCF distribution (black line) is in fact the sum of the sinusoidal signals from each internuclear distance in the molecule corresponding to two C=O bonds (green line) and the O-O internuclear distance (red line). The radial distribution function is obtained by Fourier transforming the MCF to transform from the momentum transfer space (in units of Å$^{-1}$) to radial space (in units of Å), as given by

$$g(r) = \int_0^{q_{\max}} MCF(q)\, e^{-\beta q^2} \sin(qr)\, dq, \tag{2.19}$$

where the factor $\beta$ is chosen to dampen the transform $g(r)_{q_{\max}}^{\infty}$ to zero for the limits of [$q_{\max}$, ∞]. Fig. 2.2B shows the radial distribution function for $CO_2$ where the two peaks correspond to two C=O bonds (1.16 Å) and a single O-O internuclear distance (2.32 Å). As mentioned earlier in Sec. I, the centre of each peak corresponds to the average internuclear distance whilst the FWHM of each peak corresponds to how vibrationally excited it is.

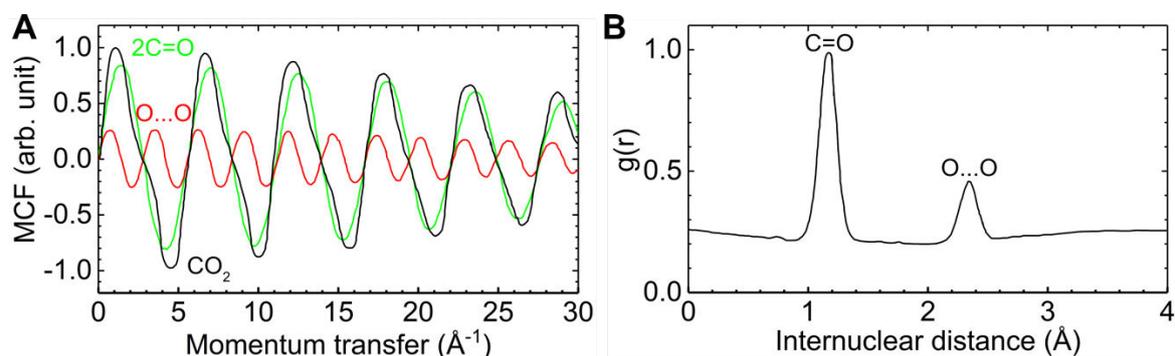

**Fig. 2.2** Molecular interference signal and structural determination of internuclear distances in $CO_2$ molecules. (A) Molecular contrast factors obtained from the inverse Fourier transform of known equilibrium internuclear distances as given by the (B) radial distribution function, $g(r)$. Here, the MCF corresponding to only the $CO_2$ (black line) as well as the individual contributions from the two C=O bonds of 1.16Å (green line) and the O-O internuclear distance of 2.32 Å (red line) are shown. Figure adapted from [Hargittai1988].

*Electron diffraction of complex polyatomic molecules*

To determine the molecular structure of more complex molecules, it is paramount that the widths of the peaks in the radial distribution function are narrow enough to adequately determine their centre positions. This requires a sufficient spatial resolution, as given by

$$\delta r_{ij} = \frac{2\pi}{q_{\max}}, \qquad (2.20)$$

where $q_{\max}$ is the maximum momentum transfer achieved which is determined by the wavelength (and thus the kinetic energy) of the scattering electron.

Additionally, alignment of molecules becomes necessary as only radial distribution functions can typically be obtained for randomly-oriented molecules without any angular information, retrieving average values of internuclear distances only. This is due to the averaging effect that exists for each two- or three-dimensional (2D or 3D) molecular orientation that "blurs" out the 2D or 3D information. For small molecules (of a few atoms), 1D information can be enough to determine the entire molecular structure particularly with the use of trigonometric and geometric considerations. However, complex polyatomic molecules usually have 2D or 3D molecular structures, and so these molecules must be aligned or some tomographic procedure must be employed to obtain their complex structures. Fig. 2.3A shows a blurred calculated structure that was extracted from simulated diffraction images of 1D-aligned molecules with an alignment cone of $30°$, whilst a well-resolved structure can be retrieved for perfectly aligned molecules in Fig. 2.3B. It should be noted though that for some molecules, it may already be sufficient to extract 3D information by only aligning one molecular axis of the molecule. In the following section, field-free electron diffraction methods will be discussed in determining the static and time-resolved structures of gas-phase molecules.

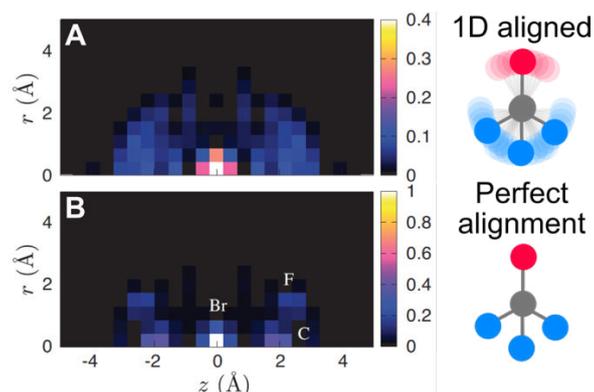

**Fig. 2.3** The two-dimensional (2D) structural retrieval of an ensemble of $CF_3Br$ molecules that were (A) laser-induced aligned with an alignment cone of $30°$ and (B) perfectly aligned along one molecular axis (*i.e.* one-dimensional alignment) using calculated X-ray diffraction data. Schematics of the molecules with their specific alignments is shown. Here, the 2D structure was retrieved by integrating the diffraction pattern along the azimuthal angle. Figure adapted from [Ho2009].

### III – Field-free electron diffraction imaging

In this section, the direct retrieval of static molecular structures in the gas-phase through conventional gas-phase electron diffraction (GED) will be discussed. This is then followed by time-resolved GED methods that are used to retrieve the time-resolved structures of gas-phase molecules through ultrafast electron diffraction (UED), in particular relativistic MeV UED.

### IIIa – Conventional gas-phase electron diffraction (GED)

Gas-phase electron diffraction (GED) can retrieve structural information of static structures with sub-Ångstrom spatial resolution [Mark1930, Hargittai1988]. In GED, a continuous beam of electrons possessing a relatively high kinetic energy (10 – 100 keV) impinges a molecular beam in a vacuum chamber. Fig. 3.1 shows the experimentally retrieved structure of diiodotetrafluoroethane ($C_2F_4I_2$) by GED. The weakly modulated calculated total electron scattering signal, $I_T$, shown in Fig 3.1A (blue solid) arises from the $I_M$. A polynomial function (red dashed) is fitted to account for the background incoherent atomic scattering signal which allows for the MCF to be extracted, as shown in Fig. 3.1B. The corresponding radial distribution function in Fig. 3.1C shows a more congested and complex distribution relative to the earlier mentioned $CO_2$ case (see Fig. 2.2). Here, the radial distribution function peaks correspond to the C-F, C-I, F-I, F-F and I-I internuclear distances in the $C_2F_4I_2$ molecule.

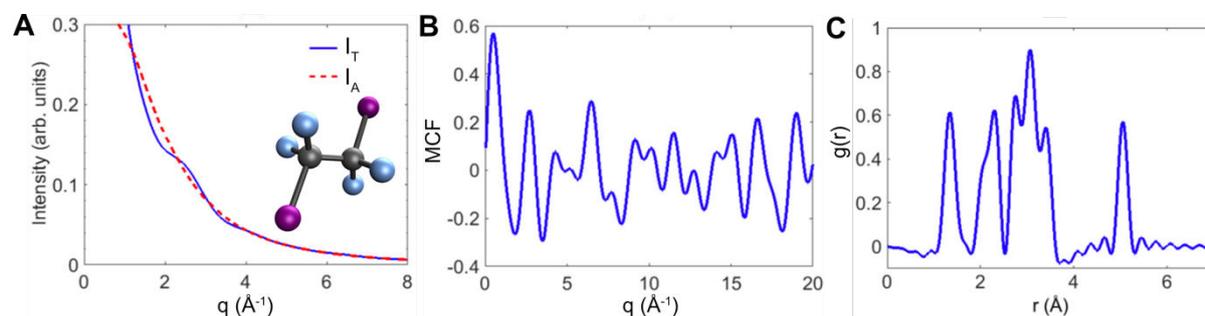

**Fig. 3.1** Structural retrieval of randomly-oriented 1,2-diiodotetrafluoroethane ($C_2F_4I_2$) molecules through calculated gas-phase electron diffraction data. (A) The total scattering signal from $C_2F_4I_2$ molecules, $I_T$. The incoherent sum of atomic scatterings, $I_A$, is extracted from the simulated $I_T$ data by fitting a background polynomial function to it. A schematic of the equilibrium ground-state $C_2F_4I_2$ molecule is shown as an inset. (B) The molecular contrast factor (MCF) is Fourier transformed to give the corresponding (C) radial distribution function function, $g(r)$ of $C_2F_4I_2$ molecules. Figure adapted from [Centurion2016].

Retrieving accurate, multi-dimensional structural information of complex gas-phase molecules will require either or both: (i) retrieval algorithms that find the calculated interference signal from quantum chemical calculations that best matches the corresponding measured signal [Sim1975, Davis1990, Klimkowski1982, Mitzel2003, Yang2018a], and (ii) as mentioned in Sec. II, alignment of molecules to give a well-defined spatial orientation. The latter can be achieved by laser-induced alignment using ultrafast optical pulses [Stapelfeldt2003, Hansen2012, Amini2017a]. As mentioned in Sec. I, recording snapshots of time-resolved molecular structures requires an imaging technique that can record a snapshot of the transient molecular structure on the nuclear (*i.e.* femtosecond) timescale. In fact, many photo-initiated chemical reactions proceed on the tens of femtoseconds to picosecond timescale. This can be achieved with sub-picosecond pulses of electron beams typically in time-resolved GED measurements commonly referred to as ultrafast electron diffraction (UED). First, a brief discussion of ultrafast electron microscopy (UEM) is given, which was the driving force in the development of ultrafast gas-phase electron diffraction studies.

### IIIb – Ultrafast electron microscopy (UEM)

Ultrafast electron microscopy (UEM) [Lobastov2005, King2005, Zewail2006, Hassan2017, Hassan2018] can image cells and material structures with the sub-nanometre spatial resolution of a transmission electron microscopy (TEM) [Reimer1997, Thomas2001]. Importantly, UEM is the time-resolved analogue of TEM, with the capability of recording images with an exposure time on the sub-picosecond timescale. Fig. 3.2A shows the typical time-resolved UEM set-up. Here, a femtosecond optical pulse generates a sub-picosecond pulsed electron beam that is subsequently accelerated and focussed onto the target sample. A second portion of the optical pulse is used as a pump pulse to photoexcite the sample. Figs. 3.2B and 3.2C show UEM diffraction patterns of amorphous carbon and single-crystal gold samples which have an excellent agreement with their corresponding TEM images shown in Figs. 3.2F and 3.2G. Interestingly, the UEM images of a rat intestinal cell shown in Fig. 3.2D and its zoom-in of a small vesicle magnified by 10 times in Fig. 3.2E show extremely detailed images of the cell structure and its sub-cellular organelles. These are again in excellent agreement to their corresponding TEM images in Figs. 3.2H and 3.2I.

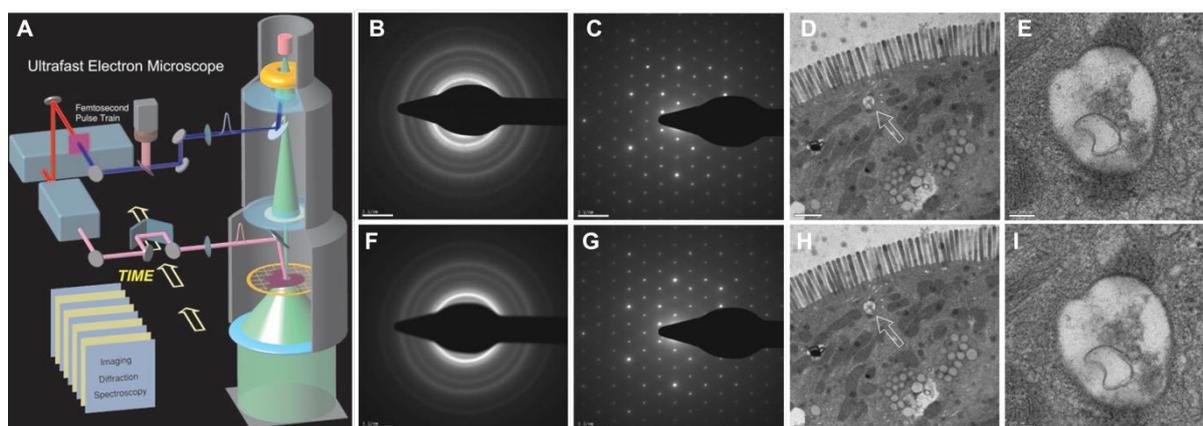

**Fig. 3.2** Ultrafast electron microscopy (UEM). (A) Schematic of a UEM experimental set-up. See text for details. (B) – (I) Images of (B)/(F) amorphous carbon, (C)/(G) single-crystal gold, (D)/(H) rat intestinal cells, and (E)/(I) a zoom-in of (D)/(H) captured with (B)-(E) UEM and (F)-(I) TEM. Figure adapted from [Lobastov2005]. Copyright (2005) National Academy of Sciences, U.S.A.

Isolated UEM pulses of down to ~30 fs (FWHM) using 200 keV electrons have been experimentally demonstrated using an optical-gate approach [Hassan2017]. Moreover, theoretical results suggest that attosecond optical gating of UEM pulses is feasible, establishing the field of "attomicroscopy" to image the motion of electrons in real-time [Hassan2017, Hassan2018].

### IIIc – Ultrafast electron diffraction (UED)

In ultrafast electron diffraction (UED) [Ihee2001, Siwick2002, Wang2003, Siwick2003, Srinivasan2003, Siwick2004, King2005, Hastings2006, Wang2006, Zewail2006, Sciaini2011, Hensley2012, Yang2016a, Yang2016b, Carbajo2016, Centurion2016, Yang2018a, Wolf2019, Shen2019, Ischenko2019], a sub-picosecond pulse of electrons is generated by illuminating a metal photocathode with a femtosecond optical pulse, as shown in Fig. 3.3. The generated electron beam is then accelerated and temporally compressed using radio frequency (RF) fields before the beam is spatially compressed and focussed onto the target sample using a magnetic lens. The incident electron beam then scatters against the nuclei and electrons of the molecules in the molecular beam jet. The transmitted scattered electron beam is then imaged using an electron detector set-up that typically consists of either a phosphor screen or a photographic plate which are coupled to a charge-coupled device (CCD) camera. It should be noted that some detector set-ups may contain a hole in the centre to allow the unscattered, highly focussed transmitted electron beam to pass through to avoid damaging the detector set-up. The total temporal resolution of time-resolved UED measurements involving a laser excitation pulse and a UED-probe pulse is typically on the order of several hundreds of femtoseconds. Achieving the required temporal resolution (*i.e.* less than 100 fs) to record well-resolved snapshots of transient molecular structures requires modifications to the UED technique. In the first instance, UED can be categorised by the kinetic energy of its electron beam, $E_\mathrm{k}^\mathrm{ele}$, into non-relativistic ($E_\mathrm{k}^\mathrm{ele}$ = tens of keV) and relativistic ($E_\mathrm{k}^\mathrm{ele}$ = 1 – 4 MeV) UED. Briefly, keV (MeV) UED can achieve a pulsed electron beam with a pulse duration of 200 – 800 fs (>150 fs).

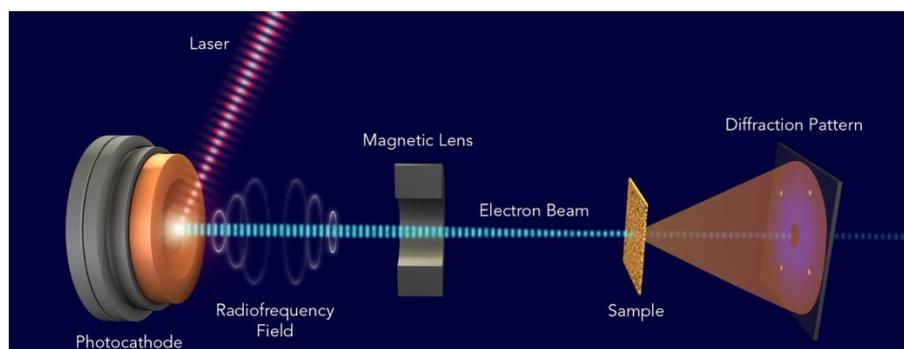

**Fig. 3.3** A schematic of a keV ultrafast electron diffraction (UED) set-up. Here, a pulsed ultraviolet laser impinges on a photocathode to generate a pulsed beam of electrons that are compressed in time using radiofrequency fields. The temporally compressed pulse is then focussed onto the sample using a magnetic lens. The diffraction patterns from elastically scattered electrons are measured using a position-sensitive detector. Figure adapted with permission from [SLAC2015].

In the following sub-sections, the temporal resolution of static non-relativistic UED measurements are discussed, followed by the total temporal resolution of time-resolved UED measurements in an optical excitation and UED probe scheme, before concluding with details of relativistic MeV UED.

*Static non-relativistic UED temporal resolution*

In static UED measurements, the pulse duration of the electron pulse, $\tau_\mathrm{e}$, at the target depends on: (i) the pulse duration of the optical laser pulse used to generate the electron beam from the photocathode, (ii) the initial velocity spread of the generated electrons, (iii) and the additional spreading due to Coulomb repulsion of electrons within the electron pulse [Centurion2016]. In fact, the latter two effects cause the electron beam to diverge in both space and time. Radiofrequency (RF) fields are employed to temporally compress the electron beam. Specifically, the time-dependent longitudinal electric field of RF fields is utilized to accelerate (decelerate) the back (front) part of the electron beam such that it arrives at the target with the shortest pulse duration possible. A magnetic lens is then used to spatially compress the electron beam from its initial several-millimetre size in order to achieve the best possible spatial resolution as it is focussed onto the sample target. It should be noted that tilting or misalignment of the magnetic lens can cause significant temporal distortions [Kreier2015].

*Time-resolved UED temporal resolution*

The total resolution of laser-pump UED-probe time-resolved measurements is given by [Centurion2016]

$$\Delta t = \sqrt{\tau_l^2 + \tau_e^2 + \tau_{GVM}^2 + \tau_{ToA}^2}, \qquad (3.1)$$

which has four contributing factors: (i) $\tau_l$ - the pulse duration of the laser pulse used to pump the sample, (ii) $\tau_e$ - the pulse duration of the electron pulse used to probe the structure, (iii) $\tau_{GVM}$ – the group velocity mismatch (GVM) between the laser and electron pulses as they traverse the target sample, and (iv) $\tau_{ToA}$ - the time-of-arrival jitter between optical and electron pulses when employing external RF fields to temporally compress or accelerate the pulsed electron beam. Today's femtosecond laser technology means that the typical pulse duration of optical pulses is less than 50 fs (FWHM), and thus, $\tau_l$ has a small influence on $\Delta t$. More significant contributions come from the three other factors in time-resolved UED measurements.

The typical pulse duration of the electron beam is around 500 fs but this can be compressed down to 200 fs using RF fields. However, this introduces a temporal jitter of 300 – 400 fs between the pulsed electron beam and optical pulse used to generate the electron beam. A time-stamping procedure overcomes this optical-electron timing jitter, achieving a $\tau_l$ of <100 fs as demonstrated in condensed matter UED studies [Gao2013]. Specifically, the time-stamping procedure uses a laser-triggered streak camera to record the arrival time of each electron pulse, allowing for the arrival times to be resorted on a shot-to-shot basis [Gao2013]. Moreover, correction of $\tau_{ToA}$ to the attosecond timescale has been proposed using a laser-cycle streaking method [Gliserin2016]. Another possibility in reducing $\tau_e$ is to minimize the propagation distance of the electron beam to the target sample, with the ability to reach down to 500 fs for a 25 keV non-relativistic pulse. However, regardless of the propagation distance, the GVM between the optical excitation pulse and the non-relativistic electron probe pulse as they traverse the target sample will contribute significantly to $\Delta t$. This is typically not a problem with optical-pump optical-probe measurements since both pulses will traverse the target sample at the speed of light in vacuum, $c$. For example, a beam of non-relativistic 25 keV electrons with a speed of $0.31c$ will take 4.3 ps to traverse a gaseous molecular beam with a diameter of 400 μm, whilst it will take 1.3 ps for the optical pulse with a speed of $c$ to travel the same distance. Thus, the GVM of roughly 3 ps will significantly contribute to the temporal blurring of the transient structures in time-resolved UED measurements. In fact, GVM not only arises from differences between the velocities of the optical and electron beams but additional temporal broadening also arises from non-collinear geometric overlap of the two pulses. Minimising the effects of GVM and temporal broadening can be achieved by (i) accelerating the pulsed electron beam to relativistic kinetic energies (*i.e.* 1 – 4 MeV) as well as (ii) using a collinear geometry [Williamson1993, Centurion2016]. Details on relativistic (MeV) UED are given in the following sub-section, with a particular focus on femtosecond-resolved UED studies.

## IIId - MeV relativistic UED

In MeV UED, pulsed electron beams can be accelerated to relativistic MeV kinetic energies using a photocathode RF gun set-up [Wang2003, King2005, Hastings2006, Wang2006, Muro'oka2011, Weathersby2015, Carbajo2016, Yang2016a, Yang2018b, Shen2019, Kim2019, Ischenko2019], which is in fact an already established technology as a bright electron source in the fields of free-electron lasers and particle acceleration [Akre2008, Gulliford2013, Terunuma2010, Yang2018b]. The relativistic nature of MeV UED helps minimize space-charge repulsion effects to subsequently generate a high bunch charge electron pulse with a temporal resolution of ~100 fs (FWHM). A schematic of a typical MeV UED set-up based on one used at Stanford Linear Accelerator Center (SLAC) [Weathersby2015, Yang2018a, Yang2018b, Shen2019] is shown in Fig. 3.4A. In this SLAC MeV UED set-up, relativistic UED beams with an electron kinetic energy of 1 – 4 MeV are typically generated through an S-band 1.6-cell photocathode RF gun consisting of two RF cavities (a half cell and a full cell), with a cross-sectional schematic shown in Fig. 3.4B [Weathersby2015, Yang2018b]. In the half-cell, a thin copper photocathode is impinged by a short (~100 fs), ultraviolet (266 nm) laser pulse generated by the frequency tripled output of an 800 nm Ti:Sapphire laser system. This generates a pulse of electrons in a similar way to that of keV UED. What is different here in the MeV UED case is that the emitted electron pulse is then accelerated out of the RF cavity to much higher (MeV) kinetic energies using the relatively high electric field (>100 MV/m) [Akre2008] in the cavity as compared to keV UED (1-10 MV/m) in order to achieve electron pulses with pulse durations close to 100 fs (FWHM) [Chatelain2012]. Just as in keV UED [Zandi2017], the MeV electron beam is temporally compressed with the ~3GHz RF field [Akre2008], before intersecting the molecular beam jet of gaseous molecules in the sample chamber. The transmitted electron beam then travels a further distance of 3.2 m towards the detector chamber, with the zero-order, undiffracted beam going through a hole located in the centre of the detector to avoid damaging the detector set-up. Just like in keV UED, the diffracted electron beam is captured on the detector and a similar data analysis to that of keV UED provides structural information of the target sample.

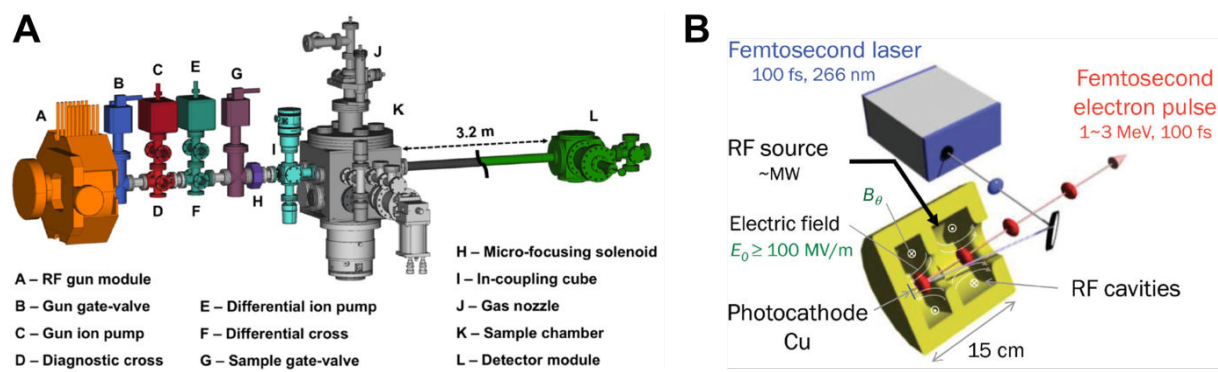

**Fig. 3.4** (A) Schematic of the MeV relativistic UED set-up, adapted from [Shen2019]. The relevant parts of the set-up are labelled accordingly in the figure and discussed in the main text. (B) Cross-section view of the radio frequency (RF) gun, adapted from [Yang2018b].

The reciprocal-space resolution of the SLAC MeV UED instrument was determined to be 0.22 Å$^{-1}$ by considering the $x$- and $y$-extent of momentum transfer ($q_x$ and $q_y$, respectively) of the (200) Bragg reflection peaks measured in the diffraction pattern of a single crystal gold sample, as shown in Fig. 3.5 [Shen2019]. A solid crystal sample is used instead of a gas-phase sample since crystal samples typically contain distinct and well-resolved diffraction spots due to their fixed lattice structures, whilst gas-phase measurements contain isotropic diffraction circles since molecules can freely vibrate and rotate.

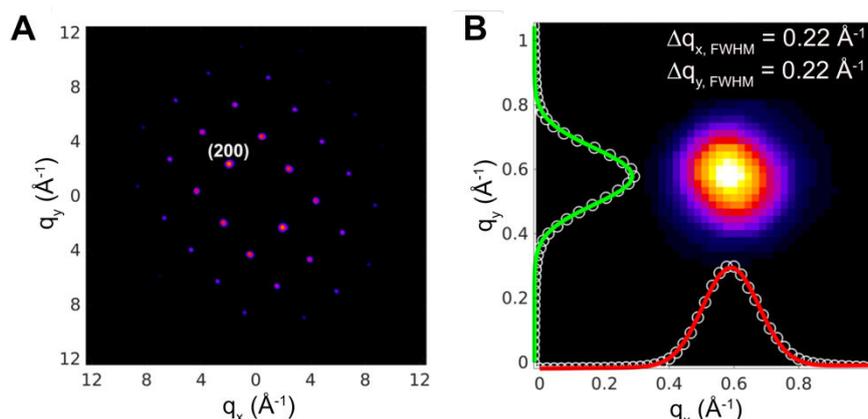

**Fig. 3.5** Reciprocal-space momentum resolution of MeV UED. (A) The electron diffraction pattern obtained from a single gold crystal sample measured using the MeV UED set-up shown in Fig. 3.4. (B) The upper limit of the reciprocal-space resolution is determined to be 0.22 Å$^{-1}$ (FWHM) from the (200) Bragg reflection peak fitted with a Gaussian distribution. Figure adapted from [Shen2019].

The spatial resolution of this MeV UED instrument was determined to be 0.63 Å through the results of static MeV UED measurements of gas-phase trifluoroiodomethane (CF$_3$I) [Shen2019], as shown in Fig. 3.6. Here, the molecular interference signal is extracted from the measured diffraction pattern (see Fig. 3.6A), highlighted through the MCF (see Fig. 3.6B), and then Fourier transformed to retrieve structural information from the corresponding radial distribution function (see Fig. 3.6C). The very good match between the calculated and measured MCF and radial distribution functions confirms that the structure of CF$_3$I can be successfully retrieved with good reciprocal-space and spatial resolution, respectively [Shen2019]. In fact, the centre of the peaks in the radial distribution function corresponding to the different internuclear distances can be found with much higher precision when the individual distributions are well-separated (*e.g.* C-F internuclear distance was measured as $1.344 \pm 0.007$ Å) [Shen2019]. Thus, the high-resolving power of MeV UED in momentum space and real space provide the capability to successfully retrieve structural information of static gas-phase molecules. The structural retrieval of transient molecular structures using time-resolved MeV UED will be discussed in the following sub-section.

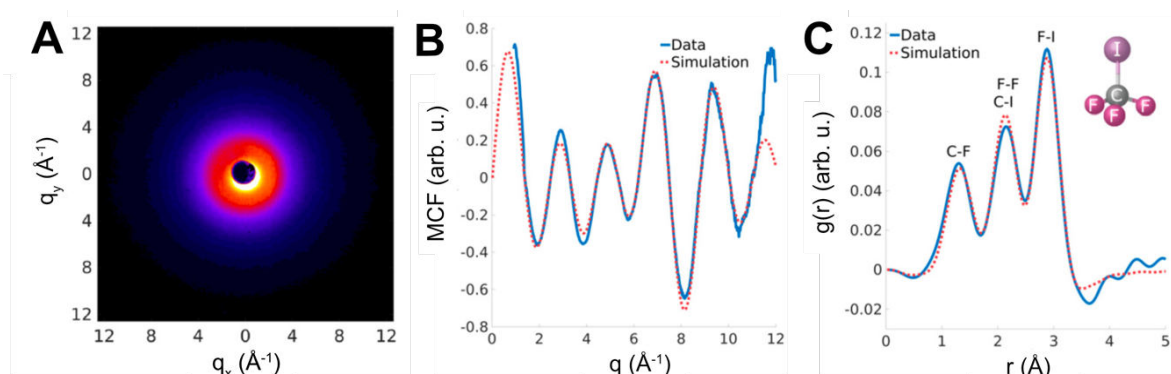

**Fig. 3.6 Static MeV UED data of trifluoroiodomethane, CF$_3$I.** (A) Measured diffraction pattern, (B) MCF as a function of momentum transfer in units of reciprocal-space, and (C) the extracted radial distribution function function, $g(r)$. Adapted from [Shen2019].

*Time-resolved pump-probe UED*

Time-resolved snapshots of a molecule undergoing a chemical reaction can be captured using the two-pulse pump-probe technique discussed in Sec. I. Here, an optical pump pulse is coupled with a MeV UED probe pulse, with multiple snapshots of the transient molecular structure recorded at various pump-probe delays. The photodissociation dynamics of CF$_3$I at 266 nm [Yang2018a, Shen2019] are used as an exemplary case to (i) determine the total temporal resolution of the SLAC MeV UED set-up, and (ii) to demonstrate ability to directly retrieve ultrafast (*e.g.* photodissociation and non-adiabatic curve crossing) dynamical information through time-resolved optical-pump UED-probe measurements.

Firstly, the total temporal resolution of the SLAC MeV UED was determined to be ~150 fs (FWHM). Fig. 3.7A shows the change in MCFs for excited molecules relative to unexcited molecules extracted from time-resolved MeV UED measurements of $CF_3I$. Here, there is a clear step-function increase and decrease (*i.e.* bleaching) in the MCF signal occurring on similar timescales. Fig. 3.7B shows the delay-dependent integrated signal corresponding to $q = 1.5 – 2.2$ Å$^{-1}$ of panel (A), which has a step-function rise time of $143 \pm 36$ fs. This is in fact the upper limit of the total temporal resolution of the MeV UED instrument which includes contributions from $\tau_e$ and $\tau_{ToA}$, with the latter determined to be ~50 fs (FWHM) using the plasma lensing method [Dantus1994, Shen2019]. In fact, the strong delay-dependent MCF changes seen in reciprocal-space at positive $\Delta t$ values result from changes in the geometric structure of the $CF_3I$ molecule following photoexcitation.

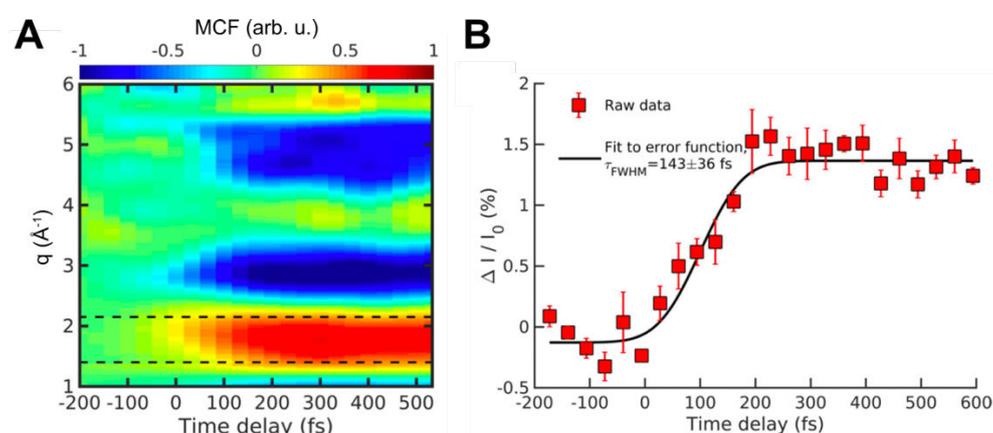

**Fig. 3.7** Temporal resolution of time-resolved MeV UED measurements of $CF_3I$ photodissociation at 266 nm using the SLAC UED set-up. (A) The change in MCF for excited molecules relative to unexcited molecules is shown as a function of momentum transfer, $q$, and pump-probe delay, $\Delta t$. (B) The delay-dependent integrated interference signal corresponding to the region of $q = 1.5 – 2.2$ Å$^{-1}$ marked as black dotted horizontal lines in panel (A). A step-function fit (black line) to the data (red squares) was applied. Figure adapted from [Shen2019].

Secondly, the photodissociation and non-adiabatic curve crossing dynamics of $CF_3I$ at 266 nm is investigated by time-resolved MeV UED [Yang2018a]. Fig. 3.8 shows the potential energy curves (PECs) involved in the one-photon photodissociation and two-photon non-adiabatic curve crossing dynamics of $CF_3I$ [Yang2018a].

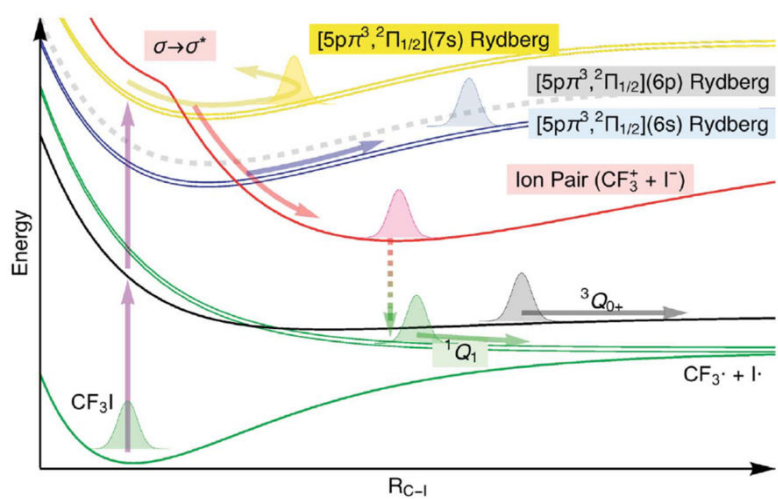

**Fig. 3.8** Potential energy curves involved in one-photon photodissociation and two-photon non-adiabatic curve crossing dynamics of CF$_3$I at 266 nm investigated with time-resolved MeV UED measurements. Figure adapted from [Yang2018a].

In the one-photon photodissociation case, the C-I bond is broken on a <200 fs timescale through a one-photon excitation to the $^3Q_0$ excited electronic state. Fig. 3.9A shows the change in the amplitude of the radial distribution function for excited molecules (*i.e.* pump-probe signal) relative to unexcited molecules (*i.e.* probe-only signal) as a function of pump-probe delay and internuclear distance. Here, two bleaching features are present corresponding to the C-I (2.1 Å) and F-I (2.9 Å) internuclear distances, with the loss of these interference signals directly related to the loss of an iodine atom following the scission of the C-I bond upon the absorption of a single 266 nm photon by CF$_3$I.

In the two-photon non-adiabatic curve crossing case, more complex molecular dynamics exists which dominates the interference signal for $\Delta t$>200 fs, as shown in Fig. 3.9B-C. The delay-dependent relative radial distribution function for two-photon excitation of CF$_3$I is shown in Fig. 3.9B. Integrating the delay-dependent signal along the initial internuclear distances of C-I (blue; 2.14 Å), F-I (black; 2.90 Å), and the distance between that of C-I and F-I (orange; 2.52 Å), as shown in Fig. 3.9C, reveals that the molecule becomes vibrationally excited at $\Delta t$>100 fs as the interference signals at 2.14 and 2.52 Å become out of phase.

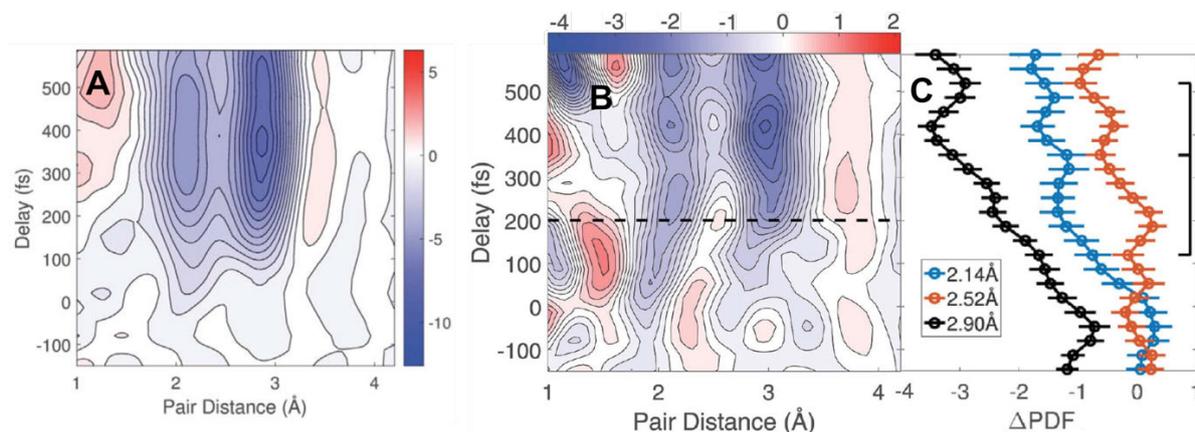

**Fig. 3.9** Time-resolved MeV UED measurements of photodissociation and non-adiabatic curve crossing dynamics in CF$_3$I at 266 nm. The change in amplitude of the radial distribution function for excited molecules relative to unexcited molecules is shown for molecules undergoing (A) photodissociation dynamics *via* one-photon excitation to the $^3Q_0$ excited electronic state and (B) non-adiabatic curve-crossing dynamics *via* two-photon excitation to the 7s Rydberg state. See Fig. 3.8 for PECs of these states. (C) The change in amplitude are shown for three internuclear distances: (i) the initial C-I internuclear distance (blue, 2.14 Å), (ii) the initial F-I internuclear distance (black, 2.90 Å), and (iii) the distance between C-I and F-I (orange, 2.52 Å). Figure adapted from [Yang2018a].

Fig. 3.10 shows the real-space reaction trajectory that can be mapped out from the detected interference signal in Fig. 3.9 using a ridge-detection algorithm which applies a background decay signal removal. In this algorithm, a background decay signal is first removed, then the ridges are identified (black dots) with the resulting trajectories established by connecting the nearby ridges (blue arrows). In fact, the experimentally obtained trajectory map (Fig. 3.10A) is confirmed by *ab initio* multiple spawning (AIMS) calculations (Fig. 3.10B) which both possess similar features in the area marked by the dashed region. The trajectory maps reveal the nuclear wave packet (NWP) dynamics involving curve crossing between the conical intersections of the initially populated 7s Rydberg state, the ion pair (IP) state of CF$_3^+$ + I$^-$, and the 6s Rydberg state. For example, the NWP bifurcates at (~2.6 Å, ~300 fs) and (2.5 Å, 450 fs), which correspond to two subsequent curve crossing events at the 7s-IP conical intersection as the 7s NWP oscillates around its minimum with a 200 fs period, with the transferred WVP then subsequently increasing in distance after curve crossing [Yang2018a].

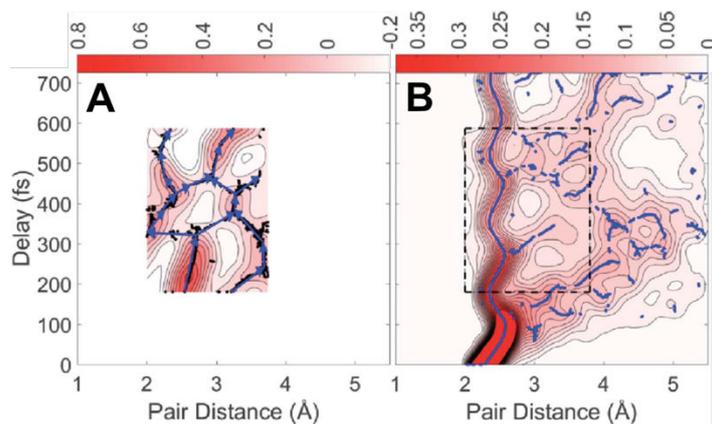

**Fig. 3.10** Nuclear wave packet trajectory map after applying the ridge-detection algorithm on time-resolved MeV UED (A) measured and (B) calculated AIMS data. The ridges are identified (black dots) and connected (blue arrow) which indicate the trajectory of the NWP as it passes near the 7s-IP and 6s-IP conical intersections. Figure adapted from [Yang2018a].

Time-resolved MeV UED measurements have also recorded molecular movies of other photo-induced dynamics, such as ring-opening in 1,3-cyclohexadiene at 266 nm [Wolf2019] and vibrational (rotational) dynamics in $I_2$ at 532 nm [Yang2016b] ($N_2$ at 800 nm [Yang2016a]).

Other approaches in minimizing the total temporal resolution of time-resolved UED measurements include using single-electron UED pulses [Lahme2014]. The single-electron nature of these pulses can theoretically produce an electron beam with few-fs to attosecond pulse duration by avoiding space-charge issues in multi-electron UED pulses. Signal-to-noise becomes prohibitive as single-electron UED pulses are "dim" electron sources relative to multiple-electron UED sources, requiring measurements to be performed at very high repetition rates (*i.e.* greater than hundreds of kHz) over long integration times (*i.e.* over many hours) with a set-up that has long-term stability. In the following section, details of field-induced electron diffraction imaging is given that are complementary to the field-free methods discussed in this section.

### IV – Field-induced electron diffraction imaging

This section explores promising ultrafast electron diffraction methods that can achieve the sub-100 fs temporal resolution required to performed time-resolved measurements of nuclear dynamics in gas-phase molecules. Here, electron diffraction is performed in the presence of an external optical field, giving rise to field-induced methods such as: laser-assisted electron diffraction (LAED), extended X-ray absorption fine structure (EXAFS), and laser-induced electron diffraction (LIED).

### IVa – Laser-assisted electron diffraction (LAED)

In laser-assisted electron diffraction (LAED) [Morimoto2014], a GED electron beam with kinetic energies of ≥1 keV is scattered against a gas-phase target in the presence of an external optical pulse. The resulting three-particle (*i.e.* electron, atom and photon) interaction can lead to the scattered electron gaining or losing kinetic energy equivalent to multiples of the photon energy, $\hbar\omega$, through a process called laser-assisted electron scattering (LAES) [Mason1993, Ehtlozky1998, Kanya2010, Kanya2011, Morimoto2015]. Fig. 4.1A shows the resulting electron diffraction patterns of carbon tetrachloride ($CCl_4$) measured with LAED [Morimoto2014]. Here, a picosecond keV electron beam impinged a molecular beam of $CCl_4$ molecules in the presence of a femtosecond near-infrared (NIR) optical pulse at varied temporal delays between the electron and optical pulses, $\Delta t$.

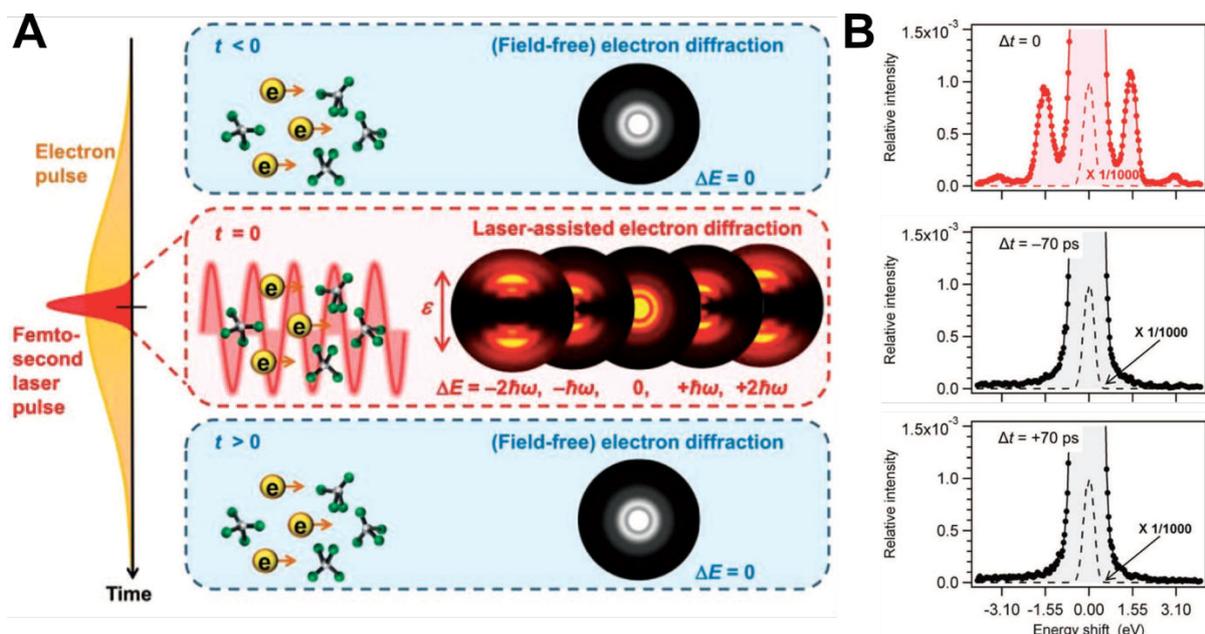

**Fig. 4.1** Laser-assisted electron diffraction of $CCl_4$. (A) A femtosecond NIR (red) pulse is overlapped with a picosecond 1 keV electron (orange) pulse to shift the energy of the scattered electrons by integers of the photon energy, $\hbar\omega$. When the optical and electron pulses are temporally overlapped (*i.e.* $\Delta t = 0$), the energy-shifted scattered electrons possess anisotropic electron diffraction patterns which contain structural information of $CCl_4$ molecules obtained within the pulse duration of the NIR pulse. Whereas when the optical and electron pulses are not temporally overlapped (*i.e.* $\Delta t = \pm 70$ ps), the unperturbed elastic electron possesses isotropic diffraction patterns. (B) The corresponding kinetic energy distribution of the scattered electrons under field-free (*i.e.* $\Delta t = \pm 70$ ps) and field-dressed (*i.e.* $\Delta t = 0$) conditions as a function of energy shift, $\Delta E$. Under field-dressed conditions, scattered electrons are detected at $\Delta E = \hbar\omega, 2\hbar\omega$. Figure adapted from [Morimoto2014].

It takes roughly 70 ps for the 1 keV electron beam to traverse the molecular beam, and so at $\Delta t = \pm 70$ ps, the scattered electrons are measured under field-free conditions. These scattered electrons possess isotropic diffraction patterns. Whereas when the two pulses are temporally overlapped (*i.e.* $\Delta t = 0$), scattered electrons are detected with an energy shift, $\Delta E$, equal to the gain or loss of one or two photons (*i.e.* $\Delta E = \pm\hbar\omega, \pm 2\hbar\omega$) as shown by the kinetic energy spectrum in Fig. 4.1B. These energy-shifted scattered electrons possess anisotropic diffraction patterns which are dependent on the scattering angle and thus molecular structure. Fig. 4.2A shows a good match between the measured (red dots) and calculated (blue line) LAED interference signal for $\Delta E = \pm\hbar\omega$ where the MCF is extracted from, as shown in Fig. 4.2B, to successfully retrieve field-free ground state structure of $CCl_4$. Thus, LAED captures structural information of gas-phase molecules within the pulse duration of the optical pulse by acting as a femtosecond optical gate in measuring diffraction patterns. Moreover, the total temporal resolution is not influenced by the picosecond pulse duration of the 1 keV electron beam due to space-charge effects nor is it affected by velocity mismatch issues.

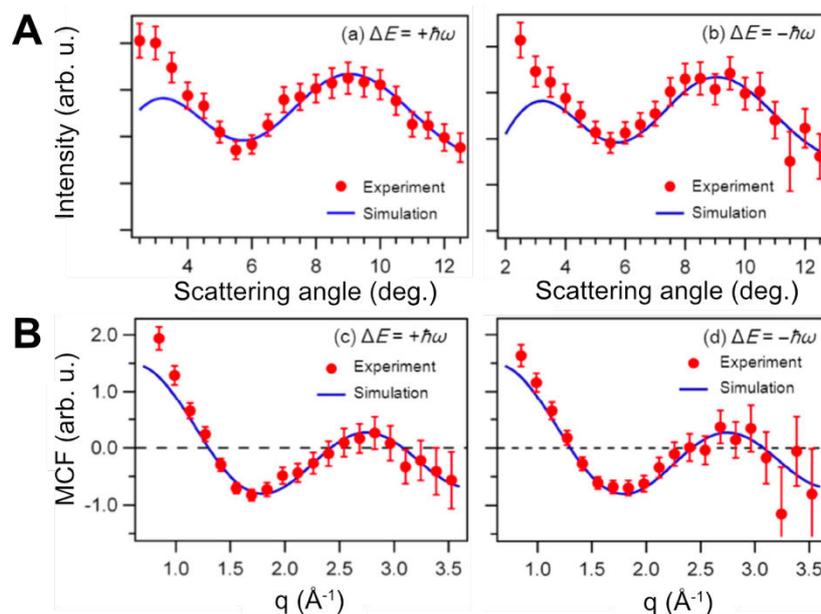

**Fig. 4.2** The experimental (red circles) and simulated (blue line) (A) interference signal and (B) MCF of CCl$_4$ obtained with LAED for $\Delta E = +\hbar\omega$ (left) and $\Delta E = -\hbar\omega$ (right) at $\Delta t = 0$ when the NIR optical field is temporally overlapped with the electron beam. Figure adapted from [Kanya2019].

Numerical calculations demonstrate the capability of LAED to reach 10-fs temporal and 0.001 Å spatial resolutions when using a THz optical field [Kanya2017, Kanya2019]. Using a THz-assisted electron diffraction (TAED) probe pulse combined with a second, separate optical excitation pulse will enable time-resolved TAED studies of gas-phase chemical reactions to be performed.

**IVb – Extended X-ray absorption fine structure (EXAFS)**

Extended X-ray absorption fine structure (EXAFS) [Rehr2000] is an intrinsic electron scattering technique where the position and behaviour of neighbouring atoms is imprinted into the electron scattering cross-section which consequently modulates the photon absorption. Thus, EXAFS provides structural information with electron resolution but using photons as the observable. In EXAFS, a significant increase in X-ray absorption is observed when the energy of the incoming photon is equal to or greater than the binding energy of a core-shell electron, commonly referred to as an absorption edge, as shown in Fig. 4.3A. The emitted core-shell electron can then scatter against neighbouring atoms, some of which can backscatter and return to the excited atom from which it was emitted from. As the backscattered electron recombines with the excited atom, it can interfere constructively or destructively with the next subsequently emitted core-electrons, as shown in Fig. 4.3A. These interferences lead to modulations in the absorption spectrum at photon energies much above the electron's binding energy corresponding to the EXAFS (blue) region as shown in Fig. 4.3B for graphite [Buades2018a]. In fact, the modulations in EXAFS signal are dependent on the position of the atoms, the identity of the atomic scatterer, the initial energy of the incoming photon, the number of scatterers, and on the strength of backscattering from the neighbouring atoms [Rehr2000]. Careful analysis of these modulations enables direct structural retrieval of the target structure, as demonstrated in Fig. 4.4 for graphite's lattice structure.

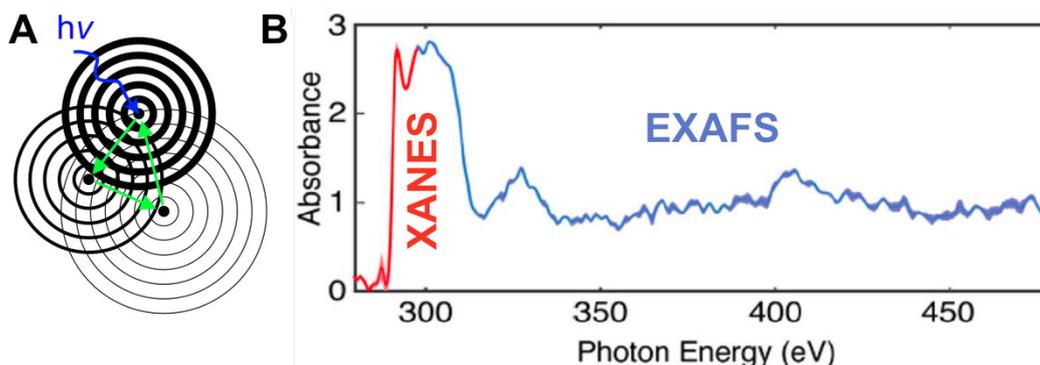

**Fig. 4.3** Extended X-ray absorption fine structure (EXAFS). (A) Schematic of EXAFS. An atom absorbs a photon with energy, $h\nu$, that is equal to or exceeds the binding energy of a core inner-shell electron. The emitted electron can backscatter against the neighbouring atoms and returned back to the initial atom it was released from, as shown by the green arrows. Constructive and destructive interferences can then occur between the backscattered electron and the next subsequently emitted electron. (B) X-ray absorption fine-structure (XAFS) spectrum of graphite as a function of photon energy obtained by high-harmonic generation (HHG). The XANES (or NEXAFS) and EXAFS portion of the XAFS spectrum are indicated in red and blue, respectively. The modulations in the EXAFS region are caused by changes in the electron scattering cross-section due to the position of the atoms in the target structure. Panels (A) and (B) were adapted from [Rehr2000] and [Buades2018a], respectively.

The modulations in the absorption spectrum due to electron-electron interferences, shown as $K\chi(k)$ in Fig. 4.4A, are extracted from the EXAFS region of the absorption spectrum (see Fig. 4.3B) following background subtraction and conversion to wavenumber space, $k$. The measured modulated absorption signal (blue circles) is then Fourier transformed to give the corresponding radial distribution function from the measured data (blue solid line) shown in Fig. 4.4B, where four peaks are clearly visible. These four peaks are identified as the electron scattering against the four nearest neighbouring carbon atoms, as shown in Fig. 4.4C. The structural retrieval is confirmed by fitting the measured data to calculated $K\chi(k)$ signal for known structures using the Artemis and Athena software packages [Ravel2005]. Fig. 4.4A in fact shows a comparison of the measured $K\chi(k)$ signal to a back-transformed fit (red solid line) in reciprocal space (Å$^{-1}$) together with the fit's uncertainty (red shaded area). Combining these calculations with the FEFF multiple scattering path simulations [Rehr2010], the contribution to the observed signal from the first four scattering paths is calculated (dashed lines in Fig. 4.4B) and the sum of the four paths (solid red line) confirms the structure retrieved by EXAFS measurements.

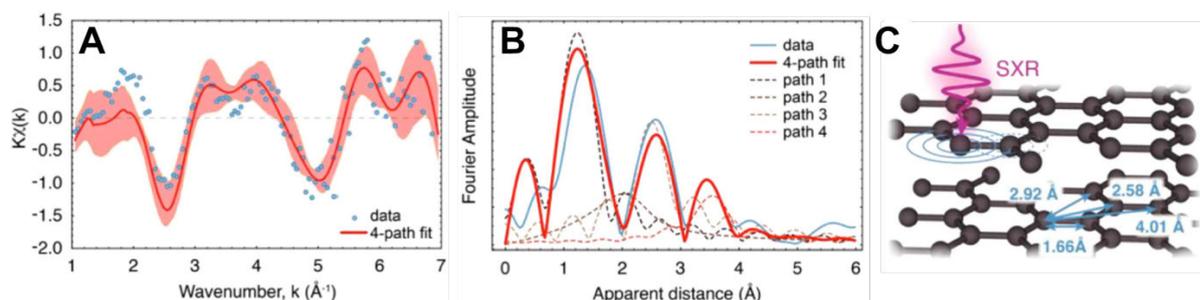

**Fig. 4.4** Structural retrieval of graphite's lattice structure via EXAFS high-harmonic generation (HHG). (A) Modulated signal in absorption spectrum due to interferences, $K\chi(k)$, as a function of electron energy in wavenumber, $k$. The back-transformed calculated (red solid line) data is fitted to the measurement (blue dots) along with the fit uncertainty (red shaded region). (B) The radial distribution function corresponding to the modulated absorption spectrum signal. The individual contributions (dashed lines) from each path of the 4-path model fit to the measured data (blue solid line) are shown along with the incoherent sum of each individual path (solid red line). (C) Schematic of the EXAFS interference signal and the measured four nearest neighbour distances in graphite. Figure adapted from [Buades2018a].

EXAFS requires significant energy range beyond the edge (*i.e.* 200 eV or more) to achieve sufficient momentum transfer. Static EXAFS is typically performed at synchrotrons [Bilderback2005, Schultz2013] but because of scanning it is always an integrated, time-averaged measurement since it is difficult to achieve a temporal resolution of less than 100 fs even with slicing sources [Schoenlein2000]. A broadband, table-top soft X-ray source can cover both the near edge (XANES/NEXAFS) and far edge (EXAFS) together in one XAFS measurement [Buades2018a]. Future possibilities include performing similar measurements at free-electron lasers (FELs) [Ackermann2007, McNeil2010, Emma2010, Chatterjee2019]. Attosecond soft X-rays [Cousin2014, Silva2015, Teichmann2016, Cousin2018] with broad enough spectra provide the possibility to perform time-resolved EXAFS measurements of photo-induced dynamics on the nuclear (*i.e.* femtosecond) and electron (*i.e.* attosecond) timescales. Specifically, high-harmonic generation (HHG) [LHuillier1993, Corkum2007, Krausz2009, Cousin2014, Calegari2016, Teichmann2016, Silva2015, Lin2018, Amini2019b] has been shown to provide table-top XUV attosecond pulses [Drescher2001, Kienberger2004, Corkum2007, Krausz2009, Sansone2010, Woerner2010, Schultz2013, Calegari2014, Kraus2015, Calegari2016, He2018]. However, core-level X-ray spectroscopy requires single attosecond pulses in the soft X-ray regime and beyond [Popmintchev2012, Cousin2014, Silav2015, Teichmann2016, Pertot2017, Cousin2018].

Fig. 4.5 shows the soft X-ray supercontinuum from an isolated attosecond pulse produced by table-top HHG which covers the entire water window (*i.e.* 200 – 500 eV) [Teichmann2016, Buades2018a, Buades2018b]. The intensity of the X-ray pulses produced by HHG are strongly dependent on the coherent addition of the harmonic fields, typically referred to as phase matching [Balcou1993, Balcou1997, Popmintchev2009, Teichmann2015, Teichmann2016, Lin2018]. Four factors typically contribute to phase-matching: (i) geometric (Gouy) dispersion – the geometric considerations of how the driver laser is focussed onto the target medium; (ii) free electron (or plasma) dispersion – the presence of ejected, free electrons changes the medium's index of refraction; (iii) neutral atom dispersion – the proportion of neutral atoms available in the medium (which is inversely correlated to the number of free electrons and free electron dispersion); and (iv) induced dipole phase – the short or long nature of the returning electron trajectory as it propagates in the laser field.

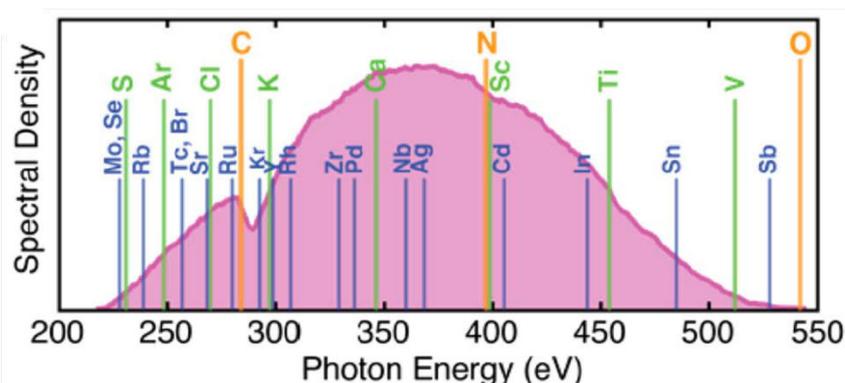

**Fig. 4.5** Spectrum showing the soft X-ray supercontinuum driven by sub-2-cycle, 12 fs 1.8 μm laser pulses used to generate an isolated attosecond pulse. Figure was adapted from [Buades2018a].

HHG can be described using the semi-classical three-step model, as shown in Fig. 4.6 [Schafer1993, Kulander1993, Corkum1993, Varro1993, Lewenstein1994]. Here, an electron wave packet (EWP) is: (i) ejected from a target structure by quantum tunnel ionization in the presence of a strong, femtosecond laser field; (ii) accelerated and returned back to the target ion by the oscillating electric field of the intense laser pulse; and (iii) recombines with the target ion, releasing its excess energy by emitting radiation as harmonics (integer multiples) of the driver laser. Especially relevant later in this chapter is the fact that the returning EWP can elastically rescatter instead of recombining with the ionic target, leading to the method of laser-induced electron diffraction. Details of strong-field quantum tunnel ionization are given in the following sub-section.

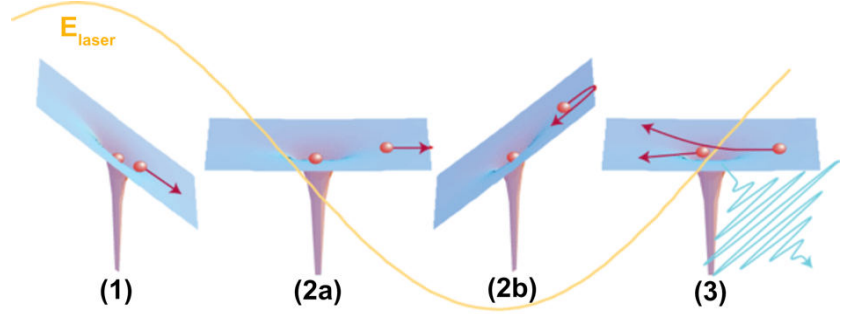

**Fig. 4.6** Semi-classical three-step model illustrating high-harmonic generation (HHG) for an atom placed in a strong field which leads to the significant perturbation of its Coulomb potential by the electric field of the laser pulse (yellow line). Here, an electron wave packet (EWP) is: (1) ejected from the target atom by quantum tunnel ionization; (2a) accelerated away from the ionized atom; before being (2b) returned back to the ionized atom as the electric field reverses sign; and (3) recombining with the parent ion, losing its excess energy gained in the field by emitting radiation as harmonics of the driver laser. It should be noted that the returning EWP in the last step can elastically (inelastically) scatter instead of recombine, leading to laser-induced electron diffraction, LIED (non-sequential double ionization, NSDI, or recollision excitation with subsequent ionization, RESI) [Amini2019b]. Figure adapted from [Corkum2007].

### IVc – Mid-infrared strong-field physics

Strong laser fields can eject an EWP from a target structure if its field strength is on the same order of magnitude as the Coulomb force of attraction between protons and electrons (*i.e.* >$10^{13}$ Wcm$^{-2}$) [Amini2019b]. This makes strong-field ionization (SFI) independent of the field's central frequency, $\omega_0$, which is in contrast to weak-field ionization based on the photoelectric effect [Einstein1905]. In the strong-field regime, several ionization regimes exist, such as: multi-photon ionization (MPI), above-threshold ionization (ATI), tunnel ionization (TI), and over-the-barrier (OTB) ionization. These SFI regimes are typically identified by the Keldysh parameter, $\gamma$, given by [Keldysh1965, Reiss2008, Eckle2008, Reiss2009, Amini2019b, Reiss2019]

$$\gamma = \sqrt{\frac{I_\text{p}}{2U_\text{p}}}, \tag{4.1}$$

which depends on the ionization potential of the target atom ($I_\text{p}$; *i.e.* the energy required to eject an electron from the target) and the ponderomotive energy of the ejected electron ($U_\text{p}$; *i.e.* the average kinetic energy of a free electron oscillating in a laser field), the latter of which is given by

$$U_p = \frac{I_0 e^2 \lambda_0^2}{8\pi^2 m_e \epsilon_0 c^3} = 9.337 \times 10^{-20} I_0 \lambda_0^2 \left[\frac{\text{eV}}{\text{Wcm}^{-2}\text{nm}^2}\right], \tag{4.2}$$

where $I_0$ and $\lambda_0$ are the peak laser intensity and central wavelength of the laser field, respectively, $e$ is the elementary charge, $m_\text{e}$ is the mass of an electron, $\epsilon_0$ is the vacuum permittivity, and $c$ is the speed of light. In relation to HHG, the maximum cut-off energy (*i.e.* the highest photon energy generated) is given by $E_\text{max} = 3.17 U_\text{p} + I_\text{p}$ [Krause1992].

However, the Keldysh parameter is used to describe tunnelling in longitudinal fields, whilst laser fields are transverse and may also require a relativistic treatment. Consequently, identifying the laser-induced SFI regimes requires the Keldysh parameter in addition to three dimensionless parameters $z, z_1, z_f$ that are obtained from the ratio of four energies $U_\mathrm{p}, w, I_\mathrm{p}, m_e c^2$ characteristic of SFI as given by [Reiss2008, Reiss2009, Reiss2019]

$$z = U_\mathrm{p}/\omega, \quad (4.3)$$
$$z_1 = 2U_\mathrm{p}/I_\mathrm{p}, \quad (4.4)$$
$$z_\mathrm{f} = 2U_\mathrm{p}/m_e c^2, \quad (4.5)$$

where $\omega$ is the frequency of the laser field. Here, $z$ provides the number of photons required to emit an electron from the target and it also gives a measure of how perturbative (*i.e.* $z \ll 1$) the ionization process is, whereas $z_1$ gives a measure of Coulomb effects that the emitted electron experiences in the presence of the laser field, whilst $z_\mathrm{f}$ provides the significance of relativistic effects [Reiss2009]. In general, these three parameters become particularly relevant with the use of long wavelength drivers where the dipole approximation [Reiss2008, Wolter2015] breaks down, magnetic field contributions to the electron dynamics become non-negligible, and relativistic effects start to become significant [Reiss2008, Reiss2019].

Fig. 4.7 illustrates the SFI processes whilst Table 4.1 summarizes their typical operating conditions and the corresponding values of the Keldysh, $z$, $z_1$ and $z_\mathrm{f}$ parameters [Reiss2009, Wolter2015, Amini2019b]. In Fig. 4.7A, the unperturbed field-free atomic Coulomb potential is shown together with the atom's ionization potential, $I_\mathrm{p}$. Placing the atom in the presence of a strong field starts to perturb the atomic Coulomb potential, causing the barrier to ionization to be lowered. In the case of MPI in Fig. 4.7B (*i.e.* $\gamma > 1$), several photons of energy $\hbar\omega$ are required to reach the ionization continuum (black arrows). It should be noted that above-threshold ionization (ATI; red arrows) is a subset of MPI where several photons above the $I_\mathrm{p}$ are absorbed, leading to a photoelectron spectrum with one or more peaks that are separated by $\hbar\omega$. When the field strength of the laser pulse is strong enough to significantly lower the barrier to ionization, an electron can either tunnel through a small barrier (*i.e.* tunnel ionization with $\gamma < 1$ – see Fig. 4.7C) or the barrier is lowered so much so that it is lower in energy than the binding energy of the electrons such that the electrons are able to "spill" out into the ionization continuum (*i.e.* over-the-barrier (OTB) ionization with $\gamma \ll 1$ – see Fig. 4.7D).

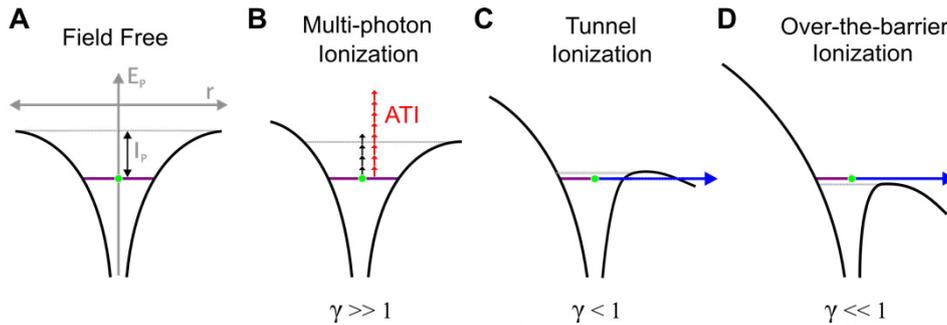

**Fig. 4.7** Ionization regimes of atom in the presence of strong fields. The atomic Coulomb potential is shown along with the atom's ionization potential, $I_\mathrm{p}$, (purple solid line) and its ionization continuum (grey dashed line). Here, the (A) unperturbed field-free and the perturbed (B) multi-photon ionization (MPI), (C) tunnel ionization and (D) over-the-barrier ionization operating regimes are illustrated. In (B), above-threshold ionization (ATI; red arrows) is a subset of MPI (black arrows) where one or more photons are absorbed above the ionization threshold. Figure adapted from [Amini2019b].

| Ionization regime | Operating condition | Keldysh parameter | $z$ | $z_1$ | $z_f$ |
|---|---|---|---|---|---|
| Single-photon ionization (SPI) | $\hbar\omega > I_p \gg U_p$ | $\gamma \ggg 1$ | $\ll 1$ | $\lll 1$ | $< 10^{-8}$ |
| Multi-photon ionization (MPI) | $I_p > \hbar\omega \gg U_p$ | $\gamma \gg 1$ | $< 1$ | $\ll 1$ | $< 10^{-6}$ |
| Above-threshold ionization (ATI) | $I_p > U_p > \hbar\omega$ | $\gamma \gg 1$ | $> 1$ | $\approx 1$ | $< 10^{-4}$ |
| Tunnel (quasistatic) ionization (TI) | $U_p > I_p > \hbar\omega$ | $\gamma < 1$ | $\gg 1$ | $> 1$ | $< 10^{-3}$ |
| Over-the-barrier (OTB) ionization | $U_p \gg I_p > \hbar\omega$ | $\gamma \ll 1$ | $\ggg 1$ | $\gg 1$ | $> 10^{-3}$ |

**Table 4.1** The typical operating conditions and the Keldysh, $z, z_1, z_f$ parameters for a variety of ionization regimes [Reiss2009, Wolter2015, Amini2019b]. Here, $\hbar\omega$ is the photon energy, $I_p$ is the ionization potential of the target, and $U_p$ is the ponderomotive energy of the emitted electron.

Performing strong-field physics (SFP) measurements deep into the quasistatic (tunnel ionization) regime enables the use of classical recollision models which are required to describe the various important of SFP phenomena that lead to HHG and LIED. In particular, these models allow the classical trajectories to be calculated and mapped to experimental features, enabling the energy and propagation times of the returning electrons to be characterized. Measurements can be performed in the quasistatic regime (*i.e.* $\gamma \approx 0.3$) by either increasing the peak laser intensity, $I_0$, or using a longer driver wavelength, $\lambda$. Fig. 4.8A shows that at a fixed peak laser intensity, $I_0$, of $1 \times 10^{14}$ Wcm$^{-2}$ typically used in near-infrared (NIR) measurements, one can perform measurements deep into the quasistatic regime by using a longer driver wavelength, $\lambda$. Whereas increasing the peak laser intensity has the disadvantage of depleting the number of ground-state molecules before the peak of the laser pulse, which is strongly dependent on the $I_p$ of the molecule. This leads to a decrease in the number of rescattered electrons generated as a result of an increase in the direct ionization of molecules on the rising edge of the laser pulse. Thus, it is important to achieve an ionization fraction of much less than unity to have sufficient neutral target density to perform rescattering measurements. To exemplify this point, Fig. 4.8B shows the fraction of molecules ionized at the peak of a six cycle laser pulse as a function of driver wavelength for two molecules at $\gamma = 0.3$: (i) naphthalene with an $I_p$ of 8 eV (which is a typical $I_p$ for most relatively large organic molecules), and (ii) acetylene with an $I_p$ of 12 eV. In the case of acetylene with a relatively high $I_p$, achieving a $\gamma = 0.3$ at 0.8 µm requires a peak intensity of $> 1 \times 10^{15}$ Wcm$^{-2}$ which leads to an ionization fraction of unity, already reaching ionization saturation. Whilst an ionization fraction of <0.01 with $I_0 < 1 \times 10^{14}$ Wcm$^{-2}$ is achievable at the longer wavelength of 3.1 µm. Interestingly, a minimum wavelength of 2.2 µm is required to achieve an appreciably low ionization fraction of <0.1 at $\gamma = 0.3$. For the case of naphthalene with a relatively low $I_p$, its ionization fraction is still unity for wavelengths up to ~2.0 µm. Whilst at 3.1 µm, its ionization fraction reaches ~0.1, providing adequate conditions to perform strong-field based rescattered measurements.

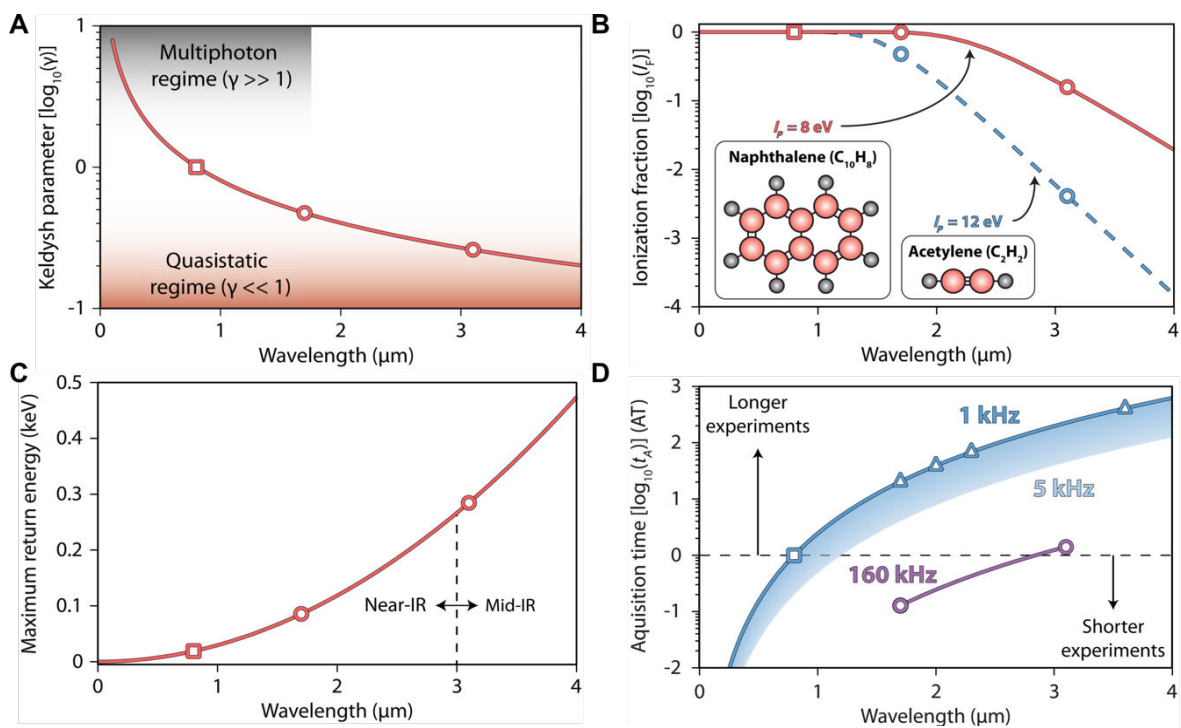

**Fig. 4.8** Mid-infrared strong-field physics. (A) Keldysh parameter, $\gamma$, as a function of a laser field's central wavelength for xenon atoms ionized with a peak intensity of $1 \times 10^{14}$ Wcm$^{-2}$. The quasistatic (red shading) and multiphoton (grey shading) regimes are indicated. (B) The fraction of molecules ionized at the peak of a six-cycle laser pulse for a variety of driver wavelengths at a constant $\gamma = 0.3$. The ionization fraction for two molecules with differing ionization potentials ($I_p$ of naphthalene, $C_{10}H_8$, is 8 eV; $I_p$ of acetylene, $C_2H_2$, is 12 eV) is shown. (C) Maximum return energy at the instance of rescattering as a function of driver wavelength at a fixed peak intensity of $1 \times 10^{14}$ Wcm$^{-2}$. (D) The acquisition time (AT) for a typical SFI measurement as a function of driver wavelength, taking into account the $\lambda^{-4}$ scaling of the cross-section for elastic electron scattering. Here, the lower the AT then the shorter the experiments will be. Typical 1 kHz (blue triangles) and 160 kHz (purple circles) systems are indicated. Figure adapted from [Wolter2015].

An additional advantage of performing MIR-SFP measurements is the significantly higher kinetic energies of the returning electrons generated in the MIR, as shown in Fig. 4.8D. This has the effect of achieving higher photon energy (electron energy) in HHG (LIED), leading to significantly improved HHG (LIED) measurements. Finally, performing SFP measurements at longer wavelengths leads to significantly lower degree of strong-field tunnel ionization, as seen in panel (B), which therefore leads to lower signal-to-noise and larger data acquisition times. Fig. 4.8D shows the typical data acquisition time required to perform strong-field physics (SFP) measurements at $\lambda = 0.8 - 4.0$ μm. Since NIR-SFP measurements are typically performed at a repetition rate of 1 kHz (blue points) over several hours or days, performing MIR-SFP measurements at the same 1 kHz repetition rate is impractical due to the more than two-orders-of-magnitude increase in acquisition time. The acquisition time of MIR-SFP measurements can be reduced to practical levels by performing MIR-SFP at the higher 160 kHz repetition rate (purple points). Thus, to perform measurements (*e.g.* LIED) under quasistatic conditions (*i.e.* $\gamma \approx 0.3$) it is important to (i) avoid ionization saturation, (ii) to operate in the mid-infrared (MIR) regime, and (iii) use a high (*i.e.* >100 kHz) repetition MIR source with practical acquisition times.

### IVd – Mid-infrared laser-induced electron diffraction

Laser-induced electron diffraction (LIED) [Zuo1996, Lein2002, Corkum2007, Meckel2008, Lin2010, Okunishi2011, Xu2012, Blaga2012, Xu2014, Pullen2015, Yu2015, Wolter2016,

Ito2016, Pullen2016, Ito2017, Amini2019a, Liu2019, Ueda2019, Amini2019b, Fuest2019, Karamatskos2019a] works similarly to HHG in the strong-field ionization and return of an electron wave packet (EWP) but differs in that the returning EWP elastically rescatters against the target ion instead of recombining. LIED can be well-described in the framework of laser-driven electron recollision [Schafer1993, Kulander1993, Corkum1993, Varro1993], with the quantum mechanical nature of the recollision process illustrated in Fig. 4.9. Here, the field-free electron density distribution corresponding to an argon $p$ atomic orbital is shown (see $t = -50$ fs). When argon is in the presence of a strong laser field, a portion of the electron density is emitted by strong field quantum tunnel ionization. This generates an attosecond EWP burst as shown by the wave fronts (blue area) whenever the intensity of the electric field is equal to or greater than the ionization intensity threshold (see $t = 0$ fs); the ionization probability is indicated as red shaded areas corresponding to specific field strengths. The oscillating electric field then returns a portion of the emitted EWP back roughly three quarters of an optical cycle later to elastically rescatter against the Ar$^+$ ion (see $t = +5$ fs). Here, the de Broglie wavelength, $\lambda_B$, and the lateral spread (1/e width), $x$, of the incoming EWP's wave fronts before elastic rescattering are shown. The resulting interference pattern after elastic rescattering contains structural information of the target which is embedded into the rescattered electron's momentum distribution that is detected with a particle spectrometer.

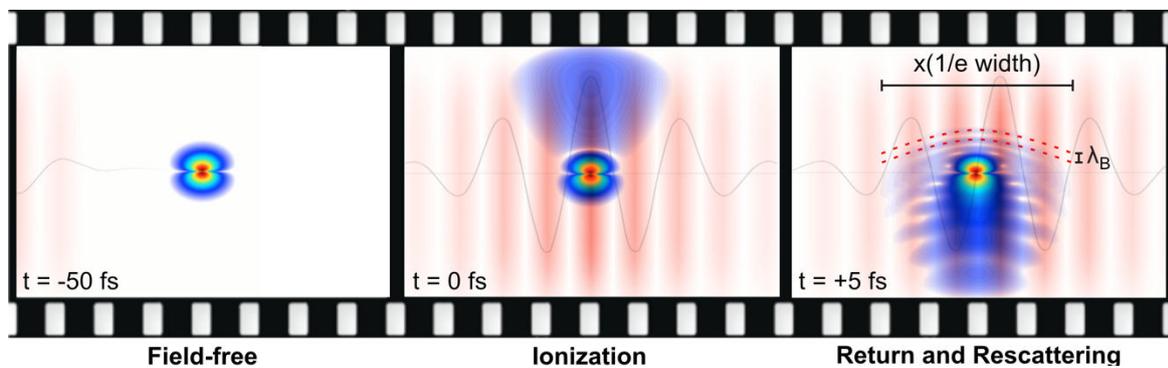

**Fig. 4.9** Quantum mechanical illustration of the laser-driven electron recollision process leading to LIED. See main text for details. Calculated using Qprop [Bauer2006].

Performing LIED measurements at longer driver wavelengths, $\lambda$, (*i.e.* transitioning from the near-infrared (NIR) to the mid-infrared (MIR) regime) provides several key advantages. Firstly, the de Broglie wavelength, $\lambda_B$, of the rescattering electron is reduced, leading to improved sub-atomic spatial resolution, as shown in Fig. 4.10 (blue line; ~0.7 Å at 3.0 μm compared to ~2.8 Å at 0.8 μm). However, the lateral extent of the rescattering EWP, $x$, becomes significantly larger at longer wavelengths (black line in Fig. 4.10), leading to a significantly reduced scattering probability, $\sigma_s$, which scales as $\lambda^{-4}$. A consequence of the reduced scattering probability at longer wavelengths is a decrease in signal-to-noise, and therefore, requiring an increase in data acquisition time. However, as discussed in Sec. IVc, the greater data acquisition time can be circumvented by using higher repetition rate MIR laser sources.

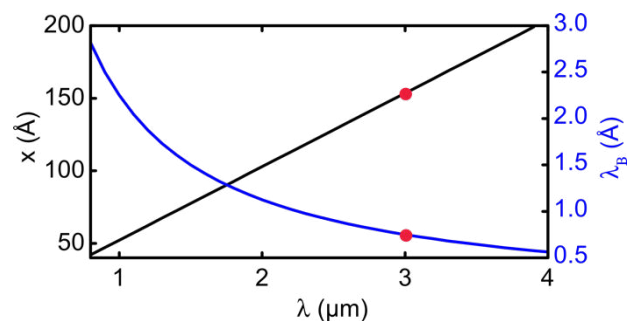

**Fig. 4.10** The de Broglie wavelength, $\lambda_B$, (blue line) of the incoming EWP and the lateral extent, $x$, (black line) of the emitted EWP by tunnel ionization as a function of driver wavelength, $\lambda$. The corresponding values at 3.0 μm is indicated by red circles. Figure adapted from [Ueda2019].

Performing LIED measurements in the MIR generates highly energetic rescattering electrons, as discussed in Sec. IVc. This leads to "hard" collisions in MIR-LIED that pinpoint the atomic locations in the target structure since the rescattering electron penetrates past the valence electron cloud (which can cause blurring effects) to directly interact with the nuclear core of each atom. Achieving such high energies at longer (i.e. MIR) wavelengths is explained by the larger distance that the emitted EWP travels for, which in turn leads to a longer time of acceleration in the field, leading to an elastically rescattered electron possessing a greater kinetic energy, $U_p$. The $\lambda^2$ scaling of the $U_p$ is very important since the maximum kinetic energy achieved with 3.1 μm is approximately 16 times larger than that of 0.8 μm, making it possible to achieve "hard" collisions at 3.1 μm.

Now let's compare the degree of scattering information that can be obtained from LIED as compared to UED. Achieving core-penetrating "hard" collisions in fact requires a sufficient momentum transfer, $q = 2k_0 \sin(\theta/2)$, of $q > 2$ Å$^{-1}$ in diffraction-based imaging experiments [Xu2010]. Fig. 4.11 shows the momentum transfer as a function of scattering angle, $\theta$, for a variety of electron kinetic energies at the instance of scattering. The typical $\theta$ ranges that CED/UED and LIED operate in are shaded in orange and blue, respectively. It is important to note that as large a momentum transfer range, $\Delta q$, is desired in order to resolve as small spatial distances as possible (see Fig. 2.2 in Sec. II). Although a similar momentum transfer range (i.e. $\Delta q = 1 - 20$ Å$^{-1}$) can be achieved for both CED/UED and LIED, the approach used to achieve high momentum transfers is quite different for both methods. In the case of CED/UED, the kinetic energy of the incoming electron beam is typically increased (e.g. see 10 keV and 50 keV distributions in Fig. 4.11). Whilst in LIED, kinetic energies of $\gtrsim 50$ eV and scattering angles of $\gtrsim 40°$ are adequate to achieve sufficient momentum transfer to pinpoint the atomic locations.

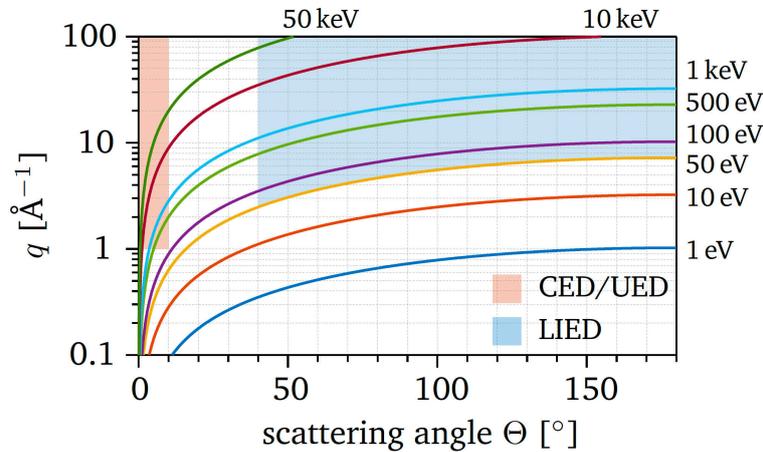

**Fig. 4.11** Momentum transfer, $q$, as a function of scattering angle, $\theta$, of the scattered electron at a variety of impact energies in eV. The typical scattering angles that CED/UED (red shading) [Ihee2001] and LIED (blue shading) [Xu2010] operate at are indicated.

To explain the scattering angle ranges that CED/UED ($\theta = 0 - 12°$) and LIED ($\theta = 40 - 180°$) measurements typically operate in, we must consider the scattering amplitudes and cross-sections as a function of scattering angle. Fig. 4.12A shows the scattering amplitude for a carbon (blue) and a hydrogen (red) atom at impact energies of 50 eV and 20 keV typically used in LIED (solid lines) and UED (dashed lines), respectively. In the case of UED, elastic scattering information is restricted to $\theta = 0 - 10°$ due to the head-on nature of forward-scattered collisions that occur at the high impact energies used in UED. Going to $\theta > 10°$ is not viable in UED since the scattering amplitude significantly decreases, as is seen for both carbon and hydrogen. Whilst in the case of LIED, scattering information can be obtained at a wide variety of $\theta$, providing greater scattering information than in UED. For example, at $\theta_r = 90°$, the scattering amplitude from LIED is approximately $10^6$ times larger than that of UED. It should also be noted that the scattering amplitude of elastic scattering is not only dependent on the impact energy and the scattering angle, $\theta_r$, of the rescattering electron but also on the identity of the atom. Typically, light atoms (*e.g.* hydrogen) have a lower scattering amplitude (and thus scattering cross-section, $\sigma$) relative to heavy atoms (*e.g.* carbon). This is clearly seen by the smaller scattering amplitude for hydrogen than carbon at all energies and scattering angles in Fig. 4.12A. To investigate whether it is possible to obtain sufficient scattering information from hydrogen relative to carbon, the ratio of the scattering cross-sections of carbon to hydrogen, $\sigma_H/\sigma_C$, is shown in Fig. 4.12B for UED at an impact energy of 25 keV (green) and LIED at 50 eV (red) and 100 eV (blue). The angular range of typical operation in UED (green shaded area) shows that $\sigma_H$ is a factor of 20 lower than $\sigma_C$, making it quite challenging to locate the position of hydrogen atoms within a molecule using UED imaging experiments. Whilst in the case of LIED, the $\sigma_H/\sigma_C$ ratio approaches within one-order-of-magnitude (*e.g.* 0.5 ratio at 50 eV, and 0.2 ratio at 100 eV), enabling LIED to resolve the atomic positions of both light and heavy atoms in a target structure across a wide variety of $\theta_r$. In particular, the appreciable scattering amplitude of hydrogen obtained with LIED makes it sensitive to locating the positions of hydrogen atoms in molecules, which other techniques (*e.g.* UED, non-covariance/coincidence CEI) are insensitive to and instead use heavier "tag" atoms such as halogens (*i.e.* iodine, bromine) to identify the molecular structure. This is particularly significant as most molecules contain hydrogen atoms, which in fact play a pivotal role in many biological and chemical processes such as hydrogen bonding in DNA and proton motion during respiration.

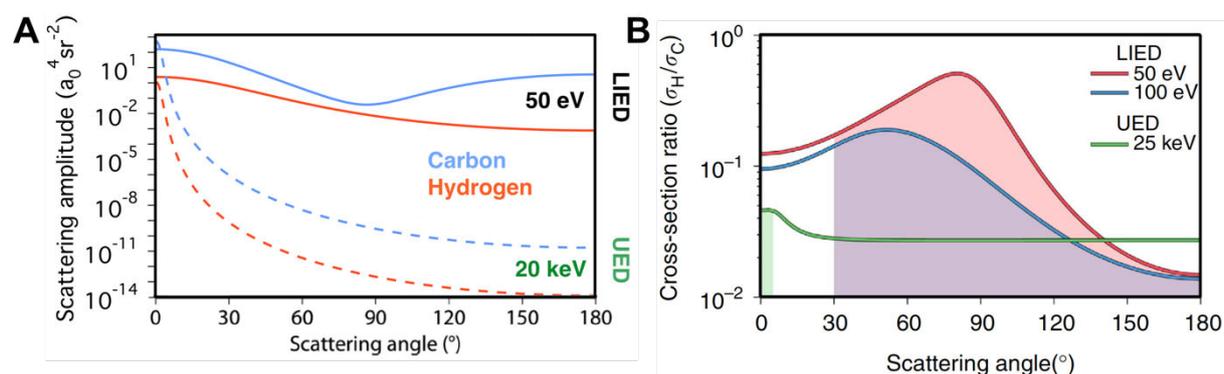

**Fig. 4.12** Comparison of LIED with UED. (A) Scattering amplitude as a function of scattering angle for elastically scattered electrons with two impinging kinetic energies corresponding to the LIED (50 eV, solid lines) and UED (20 keV, dashed lines) of carbon (blue) and hydrogen (orange) atoms. (B) The ratio of cross-sections corresponding to hydrogen and carbon, $\sigma_H/\sigma_C$, as a function of scattering angle, $\theta$, for a typical UED energy of 25 keV and LIED energies of 50 eV and 100 eV. The typical scattering angle that UED (green shading) and LIED (red and blue shading) operate at are indicated. Panel (B) was adapted from [Pullen2015].

Two important tools are required to perform MIR-LIED measurements: (i) a femtosecond MIR laser source with a repetition rate of >100 kHz, and (ii) a particle spectrometer that can simultaneously detect electrons and ions under coincidence conditions. Fig. 4.13A shows the pulse energy as a function of repetition rate for typical laser systems, with the diagonal lines indicating the average power. From this, it is clear that high average power (>10 W) is required to achieve femtosecond pulses with high pulse energies (~100 µJ) at high repetition rates (>100 kHz) in order to perform MIR-LIED measurements. For example, a 3.25 µm laser pulse with a pulse energy of 131 µJ and a pulse duration of 97 fs (FWHM; sub-9-cycle) were generated with a 21 W average power MIR optical parametric chirped pulse amplifier (OPCPA) set-up [Chalus2008, Baudisch2016, Elu2017]. These femtosecond MIR OPCPA light sources are then coupled to an electron-ion particle spectrometer, such as a reaction microscope (ReMi) [Moshammer1996, Doerner2000, Ullrich2003], a schematic of which is shown in Fig. 4.13B. Here, a molecular jet of gaseous molecules interacts with one or more ultrafast laser pulses in the interaction region to generate ions and electrons. Using a combination of electrostatic and magnetic fields from electrodes and Helmholtz coils, the electrons and ions are extracted out of the interaction region towards their own respective detector set-ups. Each detector set-up consists of two Chevron-stacked microchannel plates (MCPs) and a quad-anode delay-line anode detector. Such a set-up allows the three-dimensional momentum distribution of electrons and ions to be recorded in full kinematic coincidence (*i.e.* <0.1 event/laser shot) with <10 meV momentum resolution. Operating under coincidence conditions allows the imaging of a single, isolated molecular structure through MIR-LIED, bypassing any blurring from ensemble-averaging that CED/UED measurements may experience due to many-molecule samples. Moreover, coincidence imaging allows the disentanglement of different reaction pathways (*i.e.* dissociation, single-ionization, multiple-ionization, Coulomb explosion, NSDI, RESI, LIED) that may contribute to the detected electron-ion signal. In fact, as will be discussed later, post-processing of coincidence data will allow the signal from the process of interest (*e.g.* LIED) to be filtered and extracted from the whole dataset which in fact contains contributions from many other reaction pathways.

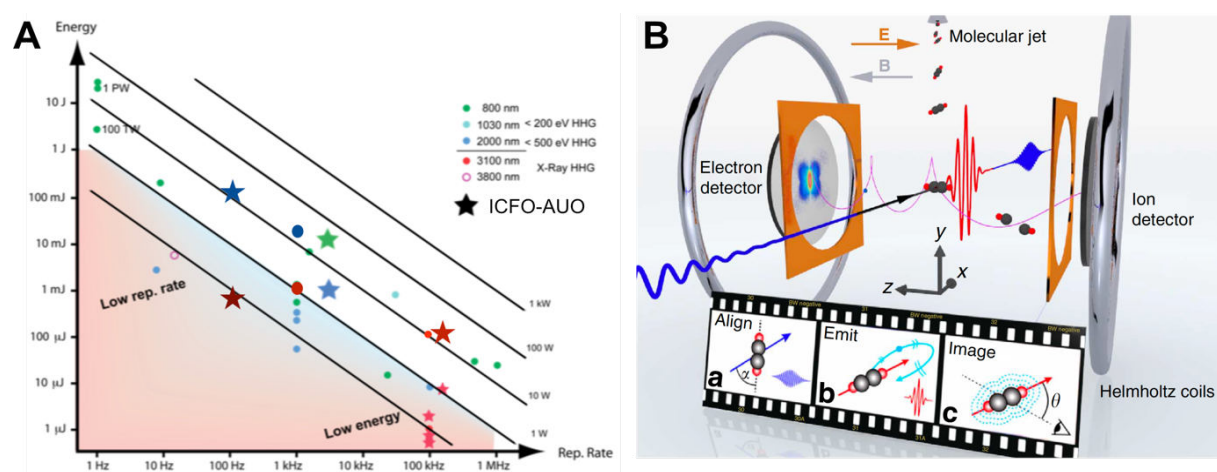

**Fig. 4.13** Tools required to perform MIR-LIED. (A) The pulse energy as a function of repetition rate of laser systems shown along with the average power required (diagonal lines). The central wavelength of laser sources is indicated by particular coloured points in the diagram, with the starred points corresponding to laser systems developed at the Attosecond and Ultrafast Optics (AUO) group in ICFO [AUO2019]. (B) Schematic of the reaction microscope coincidence particle spectrometer. See text for details. Panel (B) adapted from [Pullen2015].

Using a ReMi coupled to a MIR OPCPA laser set-up fulfils two important conditions in order to successfully perform LIED: (i) rescattering electrons must be generated with high impact energies to achieve "hard" collisions with a sufficient momentum transfer, and (ii) extraction of field-free differential cross-sections (DCSs) from a structure that is in the presence of a laser field. It is important to note that previous measurements of the DCS were restricted due to the lack of 3D momentum detection and the use of short wavelength NIR laser sources [Wolter2015]. Extension to the MIR regime has enabled the extraction of field-free DCSs of inert gas atoms [Xu2012, Wolter2015] that satisfy the quantitative rescattering model [Morishita2008, Chen2009, Lin2010]. For example, Fig. 4.14 shows the capability of MIR-LIED to extract field-free DCSs of electrons elastically scattered by Xe atoms [Wolter2015]. Here, the full 3D momentum distribution of electrons detected in coincidence with $Xe^+$ ions are shown in panel (A). From this 3D distribution, a 2D momentum distribution in cylindrical coordinates is extracted, as shown in Fig. 4.14B, which simplifies the LIED analysis and allows comparison to theory. This is achieved by applying a cartesian-to-cylindrical coordinate system transformation, followed by integration along the azimuthal angle ($\phi$), and finally the application of a Jacobian $(1\backslash|p_\perp|)$ to ensure that the correct solid-angle is used. Fig. 4.14C shows the field-free DCSs as a function of scattering angle that were extracted from the 2D distribution by integrating the electron signal in Fig. 4.14B along the circumference of a circle with a radius of the returning momentum $k_r$ that was shifted along $p_\parallel$ by the vector potential, $A_r(t)$, at the instance of rescattering, $t$. The measured field-free DCSs at the three impact energies (which is proportional to $k_r$) shown have a very good agreement with reference field-free DCSs from the NIST database (solid lines) [Jablonski2010]. Thus, this demonstrates LIED's capability to successfully retrieve field-free DCSs of a structure in the presence of a strong-field.

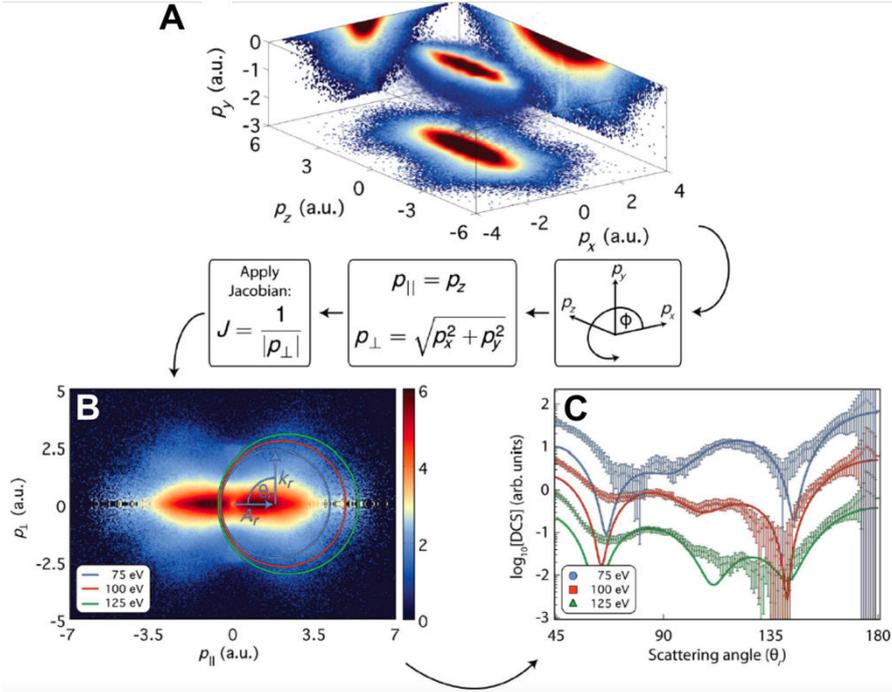

**Fig. 4.14** The capability of MIR-LIED to extract field-free DCSs from electrons elastically scattered by Xe atoms in the presence of a strong field. (A) The 3D momentum distribution of electrons detected in coincidence with Xe+ ions using a ReMi following the strong-field tunnel ionization of Xe atoms with a 70 fs (FWHM) 3.1μm laser pulse with a peak intensity of ~$7 \times 10^{13}$ Wcm$^{-2}$. (B) The 2D momentum distribution in cylindrical coordinates obtained from the 3D distribution in panel (A) is obtained following the procedure outlined in the main text. The electron signal is integrated along the circumference of three circles corresponding to three impact electron energies. Each circle has a radius corresponding to the returning momentum $k_r$ that has been shifted by the vector potential $A_r(t)$ at the instance of rescattering, $t$. (C) The field-free DCSs as a function of scattering angle, $\theta_r$, for three impact energies of 75 eV (blue circles), 100 eV (red squares) and 125 eV (green triangles) extracted from the three circles in panel (B). Reference field-free DCSs (solid lines) from a NIST database [Jablonski2010] are shown at these three impact energies. Figure adapted from [Wolter2015].

Here, we discuss two variants of LIED, called Fourier transform LIED (FT-LIED) [Pullen2016, Liu2019], also called fixed-angle broadband laser-driven electron scattering (FABLES) [Xu2014, Fuest2019], and quantitative rescattering LIED (QRS-LIED) [Morishita2008, Chen2009, Lin2010, Pullen2015, Wolter2016, Amini2019a], which are shown in Fig. 4.15. The choice of LIED variant depends on the scattering angle range of the rescattering electron investigated. In QRS-LIED, elastically rescattered electrons are investigated across a wide range of scattering angles (i.e. $\theta_r = 20 - 140°$) at a fixed return momentum, $k_r$, at the instance of rescattering. The measured 2D momentum distribution can then be compared to a simulated distribution obtained, for example, with the quantitative rescattering (QRS) theory coupled with the independent atomic-rescattering model (IAM) for known structures [Morishita2008, Chen2009, Lin2010, Kirrander2017, Yong2019]. Then, the best chi-square fit of the measured and modelled distributions enable the extraction of structural information. In FT-LIED, back-rescattered electrons (*i.e.* $\theta_r \approx 180°$) are investigated at various $k_r$, where the Fourier transform of the back-rescattered signal in the far-field provides a direct image of the object. Thus, FT-LIED can retrieve structural information without any prior structural information nor the use of structural retrieval or *ab initio* algorithms. In the following sub-section, FT-LIED will be discussed in further detail.

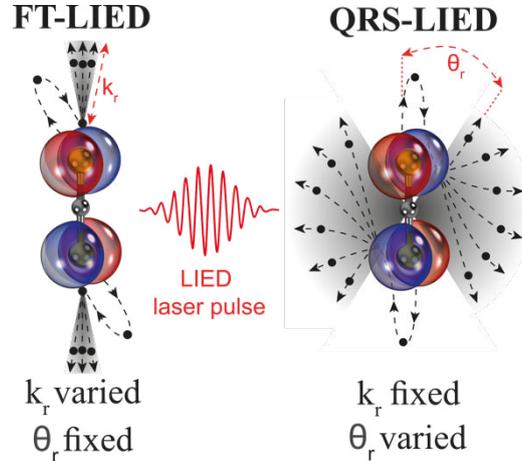

**Fig. 4.15** Variants of LIED. In FT-LIED (QRS-LIED), the returning EWP rescatters against the target structure at a varied (fixed) rescattering momentum, $k_r$, and fixed (varied) rescattering angle, $\theta_r$. Structural information is extracted from the rescattering region of interest (grey shaded area). See text for more details. Figure adapted from [Ueda2019].

*FT-LIED*

The FT-LIED analysis procedure is described here using the exemplary case of an isolated water ($H_2O$) molecule [Liu2019], as shown in Fig. 4.16. Here, the ReMi spectrometer was employed to detect, under coincidence conditions, the full 3D momentum distribution of all ions and electrons generated with a MIR laser source. The 2D momentum distribution of all electrons generated is shown in Fig. 4.16A. Here, an ion can be generated by strong-field tunnel ionization together with an electron that escapes (is returned by) the oscillating electric field of the laser pulse, referred to as a "direct" ("rescattered") electron, possessing a typical detected rescattering energy, $E_{resc}$, of $E_{resc} < 2U_p$ ($2U_p < E_{resc} < 10U_p$). In fact, ~99% of the electrons generated are direct electrons which is indicated by the rapid decline in signal due to the classical $2U_p$ cut-off energy of direct electrons (*i.e.* $P_1 = 4.7$ a.u. in Fig. 4.16) when transitioning from the direct to rescattering region. The time-of-flight (ToF) spectrum of all ions detected in the FT-LIED measurement is shown in Fig. 4.16B, with the $H_2O^+$ ToF peak dominating the spectrum. For the rest of the analysis procedure, only electrons detected in coincidence with $H_2O^+$ (see shaded area in inset of Fig. 4.16B) are considered. The field-free DCS of electrons elastically scattered off $H_2O^+$ is extracted from the 2D momentum distribution of $H_2O^+$ electrons by integrated electron counts in a block arc area similar to that indicated in Fig. 4.16A at various detected rescattering momenta, $k_{resc}$. Here, $k_{resc} = k_r + A_r$ where $k_r$ and $A_r$ are the return momentum and vector potential, respectively, at the instance of rescattering. Fig. 4.16C shows the electron counts (*i.e.* field-free DCS) for all electrons (blue dotted) and $H_2O^+$ electrons (black solid) as a function of the detected rescattering kinetic energy, $E_{resc}$. Here, the direct and rescaterring regions are clearly visible by their $2U_p$ and $10U_p$ classical cut-off energies [Krause1992] (green arrows), respectively. The inset of Fig. 4.16C highlights the importance of coincidence imaging: the $H_2O^+$ distribution has stronger modulations than in the "all" electron distribution, which in fact arises from the coherent molecular interference signal, $I_M$, that is dependent on the molecular structure (see Sec. II). Similarly to UED, the total detected interference signal, $I_T$, detected with LIED contains contributions from both the coherent $I_M$ and the incoherent sum of atomic scattering, $I_A$, the latter of which contains no structural information and contributes as a background signal.

Again, similar to UED, the molecular contrast factor (MCF) is obtained to contrast $I_M$ to $I_A$ in order to highlight the molecular interference signal of interest. To do this, $I_A$ can be either extracted from the experimental data by fitting a polynomial function (as is done here) or by calculating the $I_A$ using the IAM model (as is done typically in QRS-LIED). Fig. 4.16D shows the modulated MCF distribution as a function of momentum transfer, $q = 2k_r \cdot \sin(\theta/2)$. Similarly to UED, Fourier transforming the modulated MCF distribution in reciprocal-space (Å$^{-1}$) gives the radial distribution function in spatial coordinates (Å), as shown in Fig. 4.16E, which contains a bimodal distribution. Two individual Gaussian distributions (black dotted) are fitted to each peak in the FT spectrum (blue solid), with the centre positions of the Gaussian distributions corresponding to the average O-H and H-H internuclear distances ($R_{OH}$ and $R_{HH}$, respectively) of $R_{OH} = 1.14 \pm 0.06$ Å and $R_{HH} = 1.92 \pm 0.07$ Å, resulting in a H-O-H bond angle of $\phi_{HOH} = 115 \pm 3°$ [Liu2019]. A schematic of the retrieved $H_2O^+$ structure is shown as an inset in Fig. 4.16E, which is identified as the ground electronic state when compared to literature values [Liu2019].

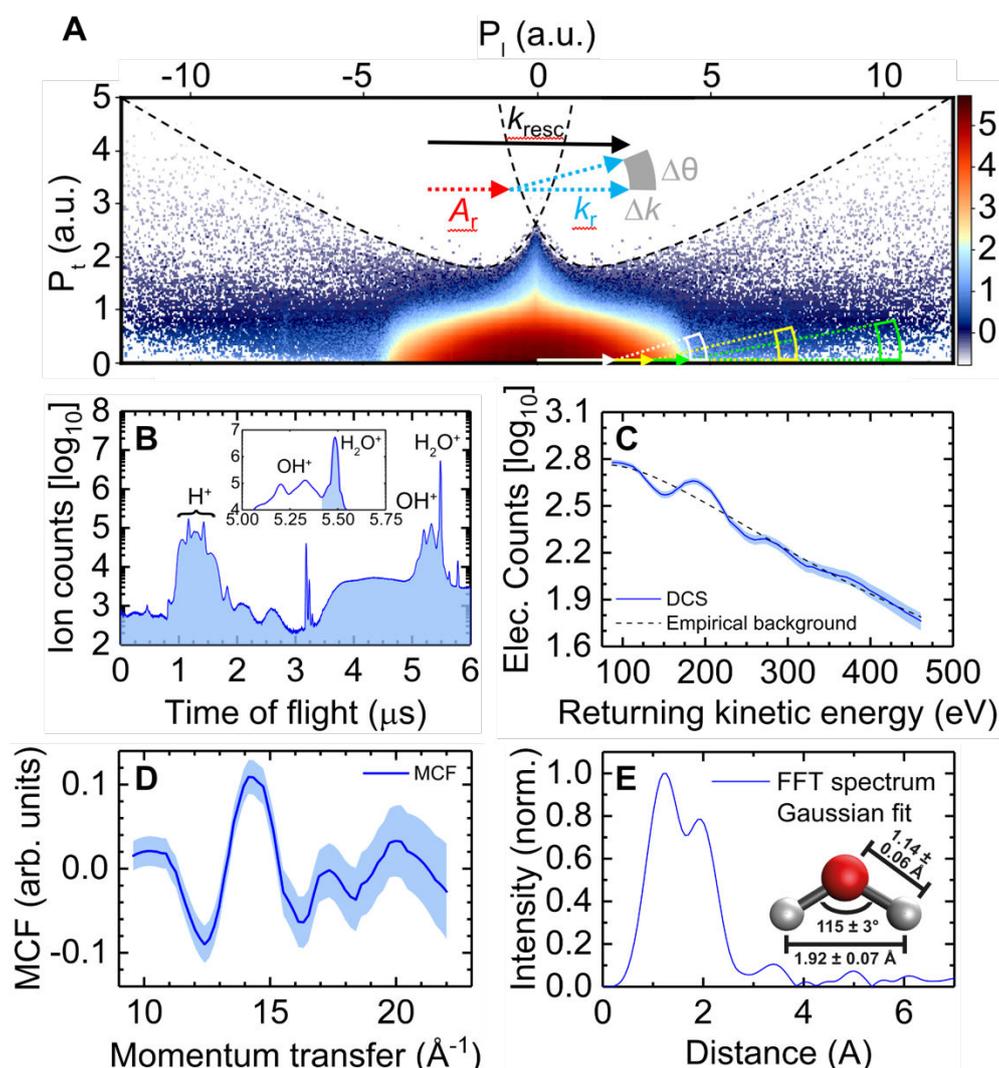

**Fig. 4.16** Data analysis procedure of $H_2O$ FT-LIED measurement. (A) The 2D momentum distribution of all electrons as a function of longitudinal ($P_l$; parallel to the laser polarization) and transverse ($P_t$; perpendicular to the laser polarization) momenta. The field-free DCS is obtained by integrating the area indicated by the block arcs at various detected rescattering momentum, $k_{resc}$. Here, $k_{resc} = k_r + A_r$, where $k_r$ and $A_r$ are the return momentum and the vector potential of the laser field at the instance of rescattering. (B) Time-of-flight spectrum of all ions generated following the ionization of isolated $H_2O$ molecules with a 97 fs (FWHM) 3.2 μm laser pulse of peak intensity, $I_0 \approx 1 \times 10^{14}$ Wcm$^{-2}$. The inset shows a zoom-in of the region-of-interest with the ToF range of $H_2O^+$ shaded. (C) Electron counts as a function of the detected rescattering kinetic energy of all electrons (blue dotted) and $H_2O^+$ electrons (black solid). The inset shows the overlap of the two distributions in the rescattering region of $2-10\ U_p$ to highlight the advantage of coincidence imaging. The $2U_p$ and $10U_p$ classical cut-off energies of the direct and rescattered electrons, respectively, are indicated by green arrows. (D) Molecular contrast factor (MCF) as a function momentum transfer. The error in the measured MCF distribution estimated *via* Poissonian statistics is shown as the blue shaded area. (E) Radial distribution function obtained from the Fast Fourier transform (FFT) of the MCF distribution in Panel (D). Here, the measured FFT distribution (blue solid) is fitted by individual Gaussian distributions (black dotted) along with the sum of the fits (black solid). A schematic of the $H_2O^+$ structure retrieved by FT-LIED is shown as an inset. Figure adapted from [Liu2019].

The coincidence capability of the ReMi spectrometer is further highlighted by the ability to directly retrieve the molecular structure of a randomly-oriented isolated molecule regardless of its highest occupied molecular orbital's (HOMO's) symmetry. For example, the HOMO of molecular oxygen ($O_2$) and acetylene ($C_2H_2$) have $\pi_g$ and $\pi_u$ symmetry, respectively, with the ionization probability of both molecules maximal along a direction other than the laser polarization (*i.e.* along 0° and 180°) as shown in Fig. 4.17A. The MCF distribution (Fig. 4.17B) for both $O_2$ and $C_2H_2$ are Fourier transformed to obtain their corresponding radial distribution functions (Fig. 4.17C), leading to the successful structural retrieval of $O_2^+$ and $C_2H_2^+$. The importance of performing LIED measurements under coincidence conditions to filter the LIED signal from all other unwanted signal is even more evident by the difficulty in retrieving the structure of $O_2^+$ under non-coincidence conditions [Xu2014].

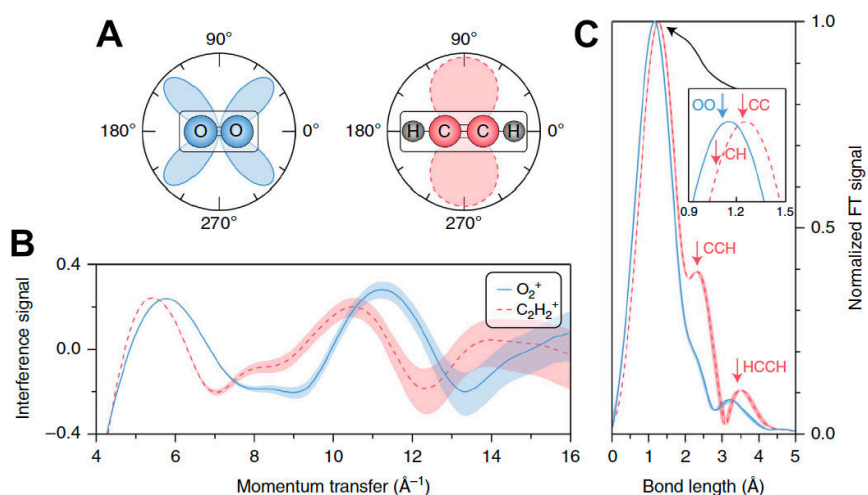

**Fig. 4.17** Structural retrieval of $O_2$ and $C_2H_2$ with FT-LIED regardless of HOMO symmetry using coincidence imaging. (A) The calculated angular-dependent ionization probability for $O_2$ (blue distributions) and $C_2H_2$ (red distributions) with $\pi_g$ and $\pi_u$ symmetries, respectively. The laser polarization is along 0° and 180°. (B) The MCF distribution for $O_2^+$ and $C_2H_2^+$ as a function of momentum transfer. (C) The radial distribution function of both molecules as a function of bond length. The respective bonds are indicated in the plot. Figure adapted from [Pullen2016].

*QRS-LIED*

As mentioned in Sec IVc, strong-field ionization (SFI) in the MIR regime enable operation deep into the quasistatic (tunnel ionization) regime, allowing the use of semi-classical models such as the QRS theory [Morishita2008, Chen2009, Lin2010] to accurately calculate SFI distributions originating from processes like LIED. Moreover, the semi-classical nature of these models enables us to include/exclude particular physical processes (*e.g.* specific types of classical trajectories of the electron born in the time-varying laser field) to accurately characterize particular aspects of the measured SFI distributions. This is especially powerful since *ab initio* time-dependent Schrödinger equation (TDSE) calculations generate the SFI distributions with all possible physical processes included (*e.g.* electron classical trajectories, EWP spreading, Coulomb focussing, angular-dependent ionization rates *etc*). Here in QRS-LIED, the experimentally measured SFI distribution is compared to a distribution calculated with QRS and IAM theories for known molecular structures in order to identify the most likely measured structure.

In QRS theory, the momentum of the photoelectron at the instance of rescattering in the laser field immediately after collision is related to the detected momentum of the rescattered electron in the laboratory frame by [Lin2018]

$$k_z = k\cos\theta = \pm A_0 \mp k_r\cos\theta_r, \quad (4.6)$$
$$k_y = k\sin\theta = k_r\sin\theta_r, \quad (4.7)$$

where $\theta_r$ and $k_r$ are the scattering angle and momentum that an electron emerges with after being elastically scattered by the target ion, whilst $\theta$ and $k$ are same quantities but in the laboratory frame, and $A(t_r)$ is the additional momentum that the rescattered electron gains from the laser field due to the field's vector potential, $A$, at the instance of rescattering, $t_r$. Here, the polarization axis of the laser field is along the $z$ axis, whilst the $y$ axis is perpendicular to it. Thus, the 2D photoelectron momentum distribution, $D(k,\theta)$, detected in the laboratory frame can be related to the momentum distribution of the incoming electron at the instance of rescattering, $W(k_r)$, and the field-free DCS, $\sigma(k_r,\theta_r)$, as given by [Chen2009, Lin2018]

$$D(k,\theta) = W(k_r) \cdot \sigma(k_r,\theta_r). \quad (4.8)$$

In fact, the magnitude of $W(k_r)$ from the returning EWP generated by strong-field tunnel ionization depends on the tunnelling ionization rate, $N(\Omega_L)$, which in turn depends on the alignment of the isolated molecule with respect to the polarization direction of the laser pulse, $\Omega_L$. Here, $N(\Omega_L)$ can be calculated using the molecular Ammosov-Delone-Kralnov (MO-ADK) theory [Lin2010, Tong2002, Zhao2010, Ammosov1986]. Thus, for a randomly-oriented isolated molecule, the expected value of $I_T$ as a function of scattering angle, $\theta$, can be written as

$$\langle I_T \rangle(\theta) = \left[\sum_i |f_i|^2 \int N(\Omega_L) d\Omega_L\right] + \left[\sum_{i \neq j} f_i f_j^* \int e^{i\vec{q}\cdot\vec{R}_{ij}} N(\Omega_L) d\Omega_L\right], \quad (4.9)$$

where $R_{ij}$ is the internuclear distance between atoms $i$ and $j$. Since $\sigma(k_r,\theta_r)$ is the same as that obtained in CED/UED measurements, the QRS model establishes a framework to obtain the field-free DCSs, and thus, structural information from electron diffraction data of rescattered LIED electrons in the presence of a strong laser field.

Fig. 4.18A shows a comparison of calculated 2D momentum maps and angle-integrated kinetic energy spectra obtained using TDSE and QRS methods for argon and xenon atoms tunnel ionized by a strong, femtosecond 0.8 μm pulse. Here, the QRS method clearly captures most of the important features that are present in the TDSE data. The difference in the features observed between the two atoms originates from the significant difference in their field-free DCSs.

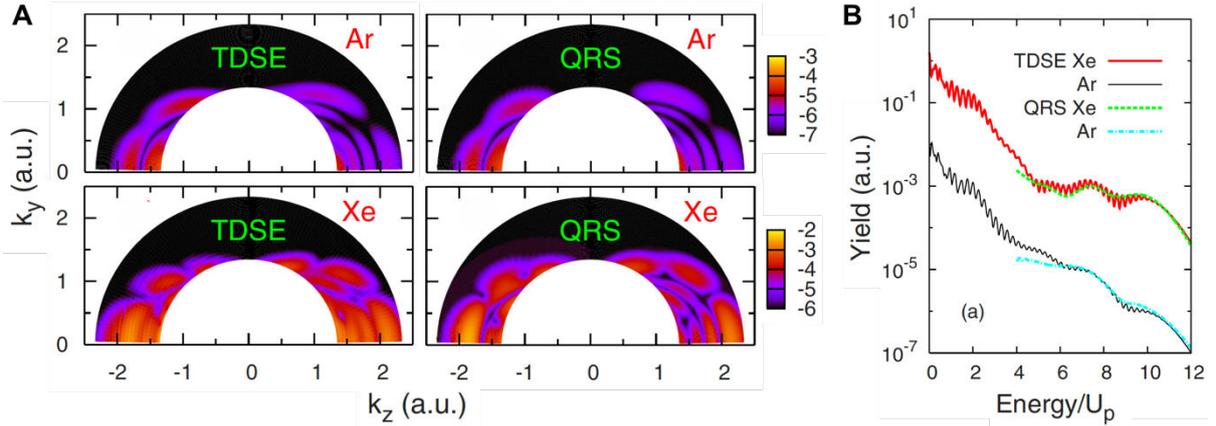

**Fig. 4.18** Comparison of electron momentum distributions calculated with TDSE and QRS methods. (A) The 2D momentum maps generated from the tunnel ionization of argon (top panels) and xenon (bottom panels) with a five-cycle 0.8 µm pulse at a peak intensity of $1 \times 10^{14}$ Wcm$^{-2}$ using the TDSE (left panels) and QRS (right panels) methods. (B) Angle-integrated kinetic energy spectra of electrons corresponding to the momentum maps in panel (A). Figure adapted from [Chen2009].

Fig. 4.19 shows QRS-LIED data for the structural retrieval of $C_2H_2^+$ [Pullen2015]. Once again, the importance of coincidence imaging is shown in Fig. 4.19A for the MCFs obtained from only $C_2H_2^+$ electrons and all electrons, with the former possessing a stronger modulated MCF distribution. To retrieve structural information from QRS-LIED measurements, a best chi-square fit of the experimentally measured MCF, $\gamma^e$, is made to a theoretical MCF, $\gamma^t$, using

$$\chi^2(R_{CC}, R_{CH}) = \sum_n [\gamma^e(k_r, \theta_n) - \gamma^t(k_r, \theta_n)]^2 \quad (4.10)$$

for a parameter space spanning, for example, the C-C and C-H internuclear distances, $R_{CC}$ and $R_{CH}$, respectively, where *n* is the index of the discretized scattering angle. The theoretical MCF that best matches the measured MCF determines the most likely geometric structure of the target ion measured by QRS-LIED. Fig. 4.19B shows the 2D $\chi^2$ map of best fit between $\gamma^e$ and $\gamma^t$ plotted in logarithmic scale as a function of $R_{CC}$ and $R_{CH}$. The most likely $C_2H_2^+$ structure measured with QRS-LIED is given at the minimum position ($\chi^2_{min}$; white dot) of the map which agrees within 10% (black vertical/horizontal lines) of the known equilibrium structure of $C_2H_2^+$ (black dot). Moreover, the MCFs corresponding to 10% changes in $R_{CC}$ and $R_{CH}$ (dashed lines) in Fig. 4.19A illustrate the sensitivity of QRS-LIED to the molecular structure.

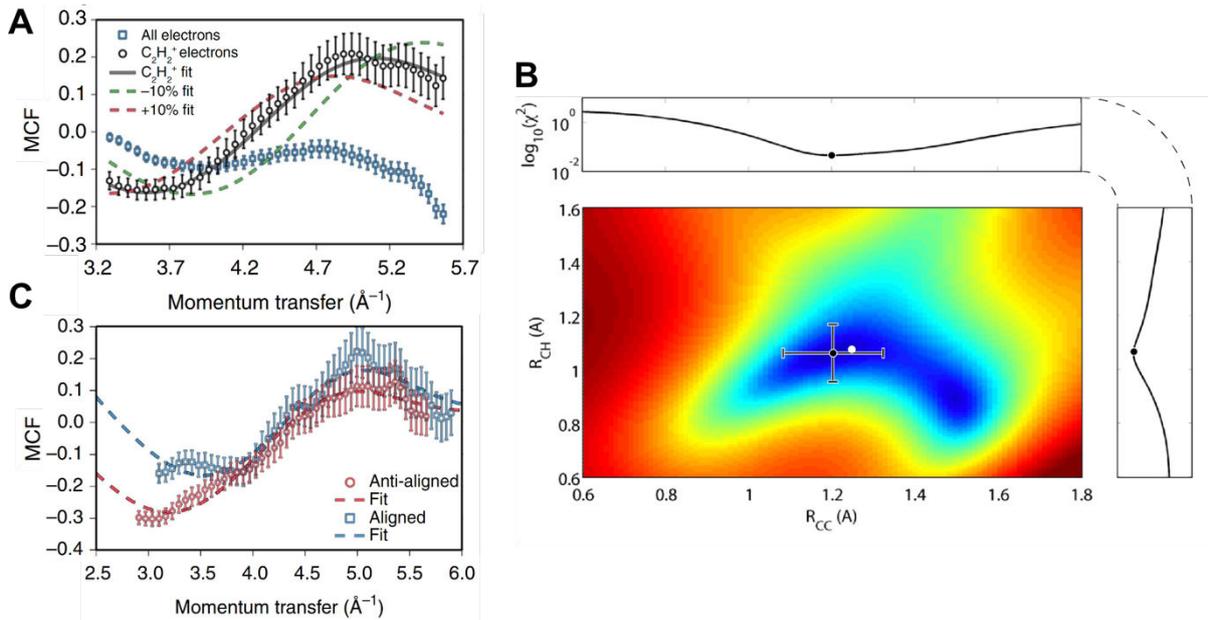

**Fig. 4.19** Structural retrieval of $C_2H_2^+$ using QRS-LIED. (A) The MCF as a function of momentum transfer for all electrons (blue squares) and $C_2H_2^+$ electrons (black circles). The theoretical MCF (grey solid line) that best fits the measured MCF distribution for the $C_2H_2^+$ electrons is shown along with the MCFs corresponding to ±10% changes in $C_2H_2^+$ structure (green and red dashed lines). (B) The 2D $\chi^2$ map showing the $\chi^2$ fitting of measured and theoretical MCFs as a function of C-C and C-H internuclear distances, $R_{CC}$ and $R_{CH}$, respectively. The best fit achieved is given by the minimum $\chi^2_{min}$ (white dot) which is in good agreement with the known equilibrium $C_2H_2^+$ structure (black dot) within ±10% error (black error bars). Cut-outs of the $\chi^2$ value shown in logarithmic scale are shown above as well as to the right of the 2D map. (C) The measured (circles and squares) and best fit (dotted lines) MCF distributions of $C_2H_2^+$ electrons from a single $C_2H_2$ molecule aligned parallel (*i.e.* aligned; blue distributions) or perpendicular (*i.e.* anti-aligned; red distributions) to the polarization axis of the LIED laser pulse. Figure adapted from [Pullen2015].

In fact, structural retrieval using QRS-LIED can be achieved from an isolated $C_2H_2$ molecule aligned parallel (*i.e.* aligned) or perpendicular (*i.e.* anti-aligned) to the polarization axis of the LIED laser pulse. This is demonstrated by the similar MCF distributions measured for aligned (blue squares) and anti-aligned (red circles) cases in Fig. 4.19C. Moreover, similar values of $R_{CC}$ and $R_{CH}$ are extracted for both alignment cases with picometre and attosecond spatio-temporal resolution, as shown in Fig. 4.20.

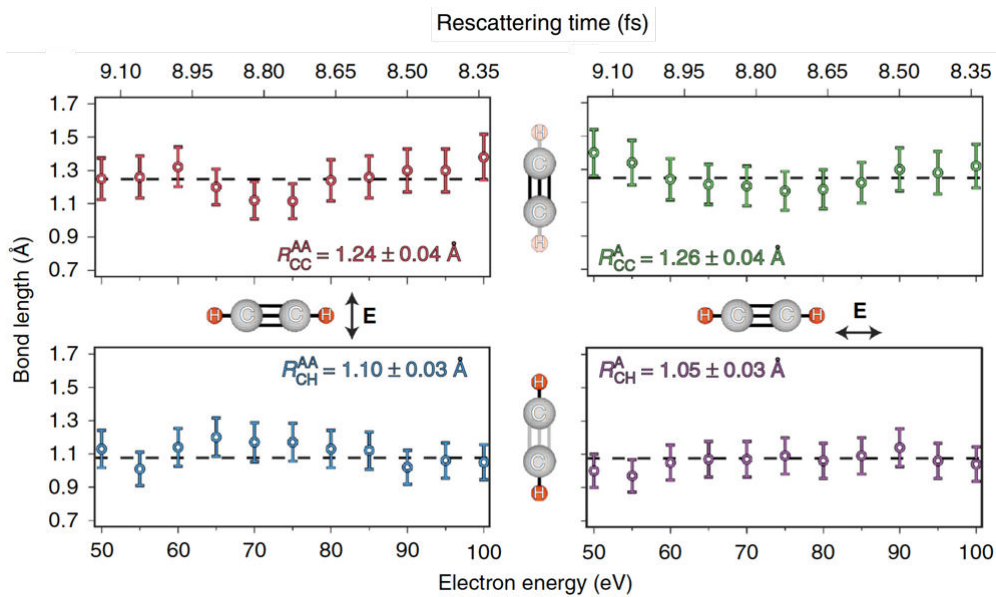

**Fig. 4.20** Direct retrieval of C-C (top panels) and C-H (bottom panels) bond lengths in $C_2H_2^+$ from an anti-aligned (AA; left panels) or aligned (A; right panels) isolated $C_2H_2$ molecule measured with QRS-LIED. The average bond length values are indicated by the horizontal dashed lines. The estimated rescattering time is also shown at the top of the figure. Figure adapted from [Pullen2015].

LIED structural imaging has retrieved the static structures of many diatomic (*e.g.* $O_2$ [Blaga2012, Xu2014, Pullen2016], $N_2$ [Xu2014]) and polyatomic (*e.g.* $CS_2$ [Amini2019a], OCS [Karamatskos2019a], $C_2H_2$ [Pullen2015, Pullen2016, Wolter2016], $C_6H_6$ [Ito2016], $C_2H_4$ [Ito2017]) molecules with picometre and attosecond spatio-temporal resolution.

*Dynamic imaging using MIR-LIED*

In this sub-section, the capability of MIR-LIED to image ultrafast structural dynamics with combined picometre and attosecond resolution for the first time is presented for (i) the deprotonation of $[C_2H_2]^{2+}$ [Wolter2016] and (ii) the linear-to-bent transition in neutral $CS_2$ [Amini2019a]. Fig. 4.21A shows the 2D momentum distribution of $[C_2H_2]^{2+}$ electrons measured with QRS-LIED for an isolated $C_2H_2$ molecule that is aligned parallel or perpendicular to the polarization axis of the LIED laser field. From these 2D momentum maps, the MCF for both alignment configurations were extracted, as shown in Fig. 4.21B, with the parallel case possessing a stronger modulated signal. The theoretical MCF that best fits the measured MCF corresponds to a retrieved C-H bond length of $2.31 \pm 0.15$ Å and $1.19 \pm 0.10$ Å ($1.94 \pm 0.10$ Å and $1.54 \pm 0.06$ Å) and a C-C bond length of $1.48 \pm 0.11$ Å ($1.38 \pm 0.06$ Å) for the parallel (perpendicular) case, with the corresponding 2D $\chi^2$ maps shown in Figs. 4.21C and 4.21D for both alignment cases. In fact, the retrieved C-H and C-C bond lengths are significantly larger than the equilibrium values of $1.06$ Å and $1.20$ Å, respectively [Haynes2013], with the C-H bond length more than doubling relative to its equilibrium value. Consequently, this was the first visualization of a deprotonation reaction. The deprotonation of $[C_2H_2]^{2+}$ in the presence of a strong field can be explained by considering the field-free (red lines) and field-dressed (green and blue lines) potential energy curves (PECs) shown in Figs. 4.21E and 4.21F. In the parallel case (Fig. 4.21E), the rising edge of the laser field before ionization leads to the PECs becoming more bound (see green PECs), causing the stretching of the C-H bonds. The moment the electric field reverses its direction (*i.e.* at the instance of ionization to initiate the LIED process), the PECs become strongly dissociative (see blue PECs). Thus, the LIED electron rescatters against a dissociating $[C_2H_2]^{2+}$ ion at the time of rescattering three-quarters of an optical cycle later (*i.e.* 7-9 fs after ionization) where its C-H bond length has doubled [Wolter2016]. In the perpendicular case (Fig. 4.21F), field-dressing the PECs does not change them significantly, with the larger C-H bond length thought to arise from the C-H bond vibrating until its eventual dissociation [Wolter2016].

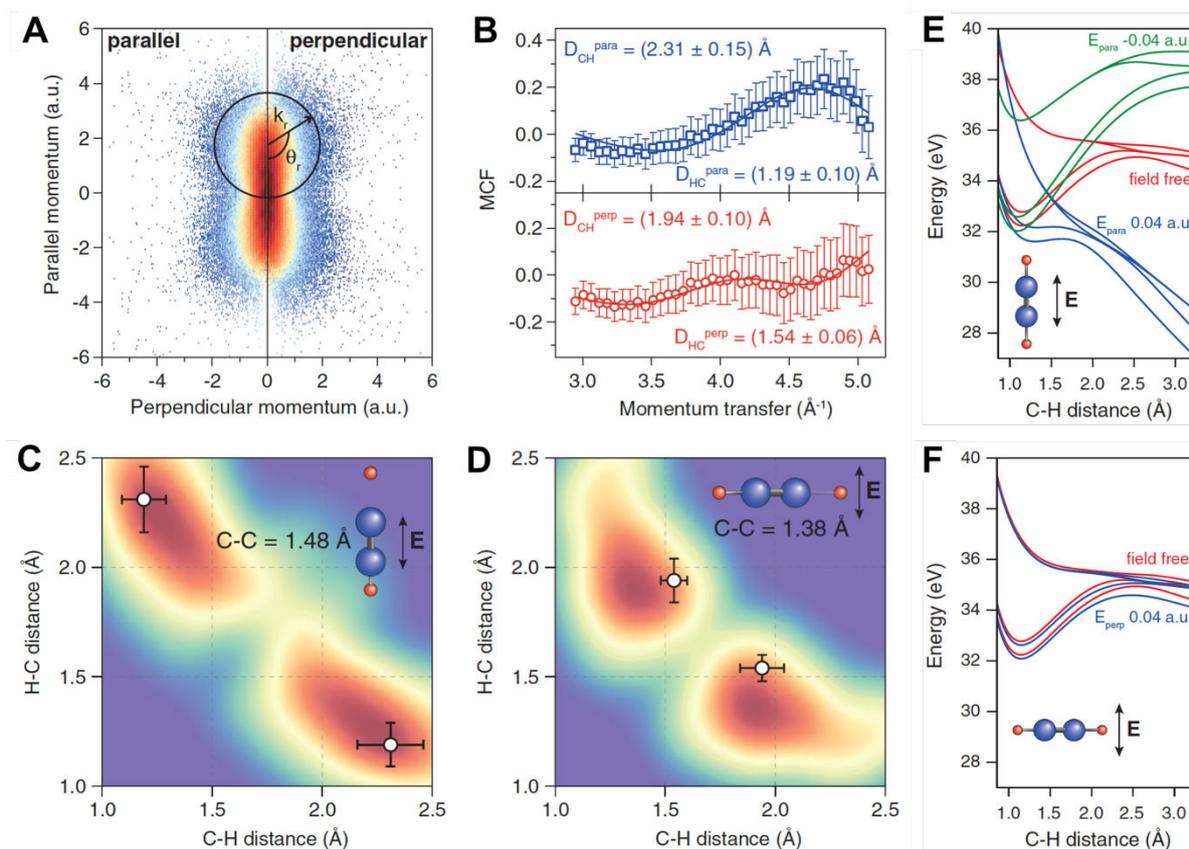

**Fig. 4.21** Imaging deprotonation reaction in [$C_2H_2$]$^{2+}$ using QRS-LIED. (A) The 2D momentum distribution of [$C_2H_2$]$^{2+}$ electrons generated from an isolated $C_2H_2$ molecule aligned parallel or perpendicular to the orientation axis of the LIED laser pulse. (B) The measured MCF distributions for parallel (blue squares) and perpendicular (red circles) alignment cases together with the theoretical MCF (solid lines) that best matches the measured distribution. The corresponding 2D $\chi^2$ maps for the (C) parallel and (D) perpendicular cases. The best fit between the measured and theoretical MCFs is given by the white point together with its corresponding error bars. The field-free (red curves) and field-dressed (green and blue curves) potential energy curves (PECs) are shown for the (E) parallel and (F) perpendicular cases. Figure was adapted from [Wolter2016].

In fact, we demonstrate that using LIED with field-dressed conditions presents no impediment to retrieving field-free neutral dynamics. Fig. 4.22 shows the structural retrieval of an isolated field-dressed carbon disulphide ($CS_2$) molecule using QRS-LIED. Here, a strongly symmetrically stretched and bent $CS_2^+$ was measured in the presence of a strong-field from the retrieved C-S internuclear distance, $R_{CS}$, and S-C-S bond angle, $\Phi_{SCS}$, shown in Fig. 4.22A.

Since field-free neutral $CS_2$ in the ground electronic state has a linear structure, a linear-to-bent transition must have occurred in the presence of a strong-field. State-of-the-art quantum dynamical wave packet calculations are shown in Fig. 4.22B which confirm that a linear-to-bent transition occurs due to the Renner-Teller (RT) effect in field-dressed neutral $CS_2$. The RT effect is the splitting of degenerate electronic states in a linear molecule to non-degenerate states as the molecule moves to a non-linear structure. Here, wave packet calculations for neutral $CS_2$ (left panels) and the $CS_2^+$ cation (right panels) are shown as a function of $R_{CS}$ (top panels) and $\Phi_{SCS}$ (bottom panels), with population in the excited electronic state shown. The initial conditions used in the neutral $CS_2$ calculation were ($R_{CS}, \Phi_{SCS}$) = (1.55 Å, 180°) corresponding to the ground electronic state of $CS_2$. In this calculation, population transfer from the $\widetilde{X}^1\Sigma_g^+$ ground electronic state to the $\widetilde{B}^1B_2$ excited electronic becomes dipole allowed (blue oval) only when the neutral $CS_2$ molecule starts to slightly bend by ~10°, which is enabled by the Renner-Teller effect in the presence of a strong field. The $CS_2$ molecule then becomes further bent at roughly the peak of the laser pulse (*i.e.* time = 0 fs). This bent and stretched neutral CS$_2$ structure in the $\widetilde{B}^1B_2$ excited state is subsequently used as the initial condition for the CS$_2^+$ cation calculation. The CS$_2^+$ cation in fact does not change in structure significantly during the rescattering time (*i.e.* 7-9 fs after ionization). A schematic of the linear-to-bent transition is illustrated in Fig. 4.22C, which shows the linear field-free CS$_2$ molecule (see $t_0$) bending on the rising edge of the pulse (see $t_p$) before it is ionized (see $t_i$) a bent and symmetrically stretched CS$_2^+$ is imaged around the peak of the pulse (see $t_r$).

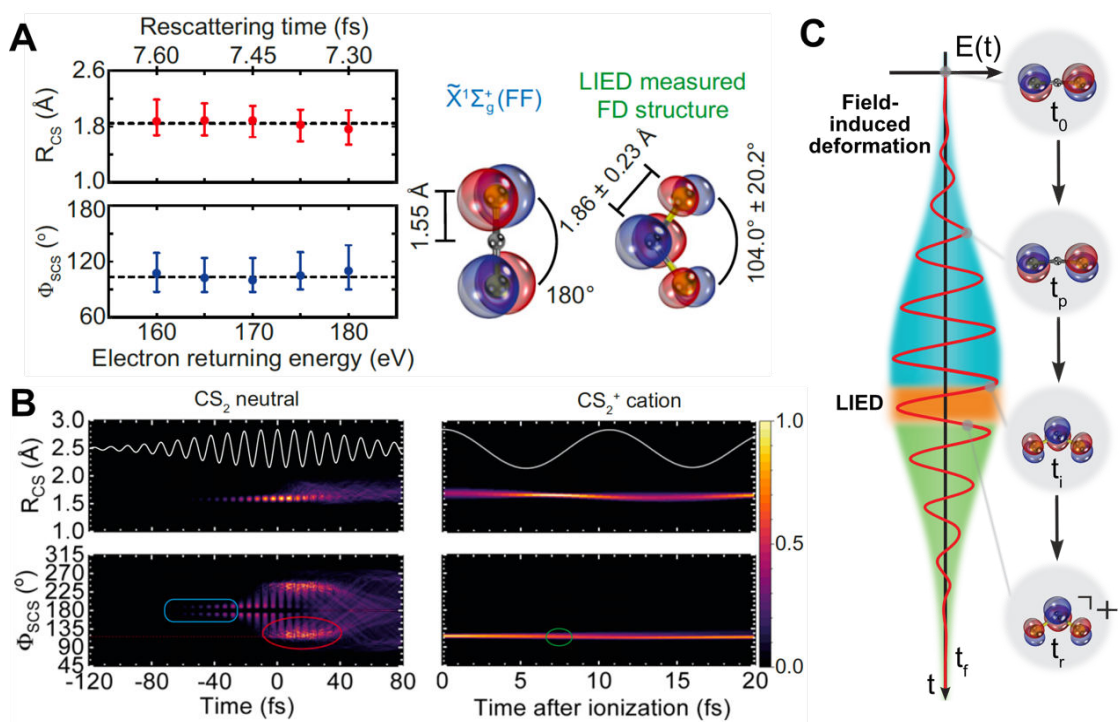

**Fig. 4.22** Imaging the ultrafast linear-to-bent neutral CS$_2$ using QRS-LIED. (A) The measured C-S internuclear distance, $R_{CS}$, and the S-C-S bond angle, $\Phi_{SCS}$, as a function of electron return energy. The geometries of field-free neutral CS$_2$ in its ground electronic state, $\widetilde{X}^1\Sigma_g^+$, and LIED measured field-dressed (FD) structures are shown. (B) Quantum dynamical wave packet calculations as a function of $R_{CS}$ and $\Phi_{SCS}$ are shown for neutral CS$_2$ during the rising edge of the field and the CS$_2^+$ cation after the neutral molecule was ionized. See main text for further details. (C) A schematic of the linear-to-bent transition of field-dressed CS$_2$ measured with QRS-LIED. Figure was adapted from [Amini2019a]. Copyright (2019) National Academy of Sciences.

As demonstrated so far, LIED has the capability of directly retrieving structural information with sub-atomic picometre and attosecond spatio-temporal resolution. Future prospects of LIED include utilizing LIED as a probe-pulse in a time-resolved optical-pump-pulse LIED-probe set-up, in which the MIR LIED pulse images the transient molecular structure at various different pump-probe delays. Consequently, an attosecond "molecular movie" of the chemical reaction can be captured to reveal intricate details of the reaction mechanism. Recording such a molecular movie with few-fs to attosecond temporal resolution of large complex molecules in the gas-phase requires several current and future challenges to be addressed, details of which are given in the following section.

## V – Future outlook and challenges

Several current and future challenges exist in performing time-resolved electron diffraction measurements on large, complex molecules, for example: preparing gas-phase molecular jets of complex neutral molecules, developing few-fs tuneable ultraviolet (UV) pump pulse, and the need for advanced retrieval algorithms.

Firstly, larger, more complex molecules typically exist as liquids and solids at room temperature (~25 °C) and pressure (1 atm), making it challenging to prepare a gas-phase molecular jet of neutral ground-state species to perform gas-phase measurements. A molecule placed in a closed vessel will have a certain pressure of its vapour in constant thermodynamic equilibrium with its condensed (*i.e.* liquid, solid) forms at a given temperature, referred to as the vapour pressure. The vapour pressure is in fact dependent on three factors: the temperature of the system, the surface area of interaction, and intermolecular forces. Fig. 5.1 shows that the vapour pressure of water increases as the temperature of the system is raised. A carrier gas (*e.g.* helium) can be used to pick up gaseous molecules by utilizing the intermolecular force of attraction between the carrier gas and the vapourised sample molecules. Finally, a larger surface area of the closed vessel will lead to a higher evaporation rate of the sample.

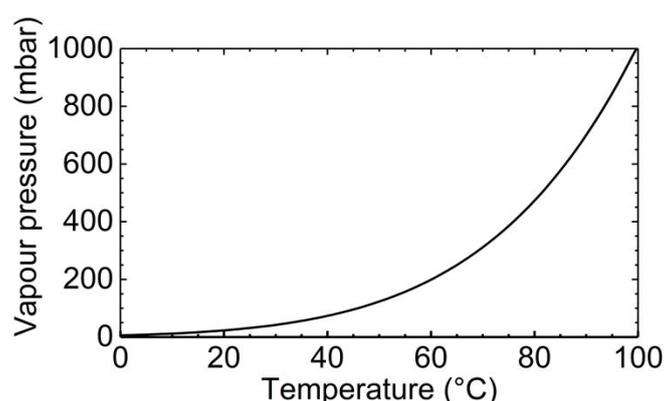

**Fig 5.1** Vapour pressure of water as a function of temperature. Data used from [Speight2017].

Secondly, generating a full attosecond resolved "molecular movie" of a chemical reaction in a time-resolved pump-probe measurement will require a pump pulse with a pulse duration (energy) of sub-fs (>0.5 µJ). Moreover, since most chemical processes are initiated by ultraviolet (UV; 200 – 400 nm) and vacuum-UV (VUV; 100 – 200 nm) radiation, pump pulses with a tuneable central wavelength in these ranges would be required. Additionally, UV single-photon excitation provides a clean path to populating a specific excited electronic state, but excitation fractions are smaller than for strong-field excitation. However, this is a universal issue and not limited to LIED or UED. Traditional approaches include generating non-tuneable 266 nm UV pulses using non-linear crystals to sum frequency mix 800 nm fundamental output from a Ti:Sapphire femtosecond laser with its second harmonic 400 nm. However, the pulse duration of the generated 266 nm pulse typically spans more than 50 fs and is not a suitable for performing attosecond-resolved UV-pump LIED-probe measurements. The third harmonic of 800 nm radiation in a gas cell has been demonstrated to generate sub-2-fs, 0.15 µJ deep ultraviolet radiation [Galli2019]. An alternative approach would be to use hollow capillary fibres (HCFs) [Travers2019, Brahms2019] and photonic crystal fibres (PCFs) [Russell2014].

HCFs have been shown to generate sub-3 fs, tuneable UV-to-VUV pulses with pulse energies of 1-10 µJ through optical soliton dynamics in HCFs. Fig. 5.2 A shows a schematic of a gas-filled HCF pumped by a 10 fs 800 nm pulse with up to 1 mJ pulse energy, with the resulting soliton dynamics combining both nonlinearity and dispersion to generate self-compressed VUV pulses. The spectrum of the self-compressed pulse was obtained using a grating. Fig. 5.2B demonstrates that the input pulse can be self-compressed from 10 fs down to 1.2 fs using an input pulse energy of 337 µJ. In fact, the FWHM of the square of its electric field is 412 as (see green distribution), with the soliton self-compressing to an optical attosecond pulse [Travers2019]. Fig. 5.2C demonstrates that a tuneable spectrum can be generated with the central wavelength varied from 122 nm to 350 nm when changing the helium input gas pressure from 230 mbar to 4 bar. Moreover, sub-1 fs tuneable UV-to-VUV pulses are generated with pulse energies of between 1 – 10 µJ as shown in Fig. 5.2D.

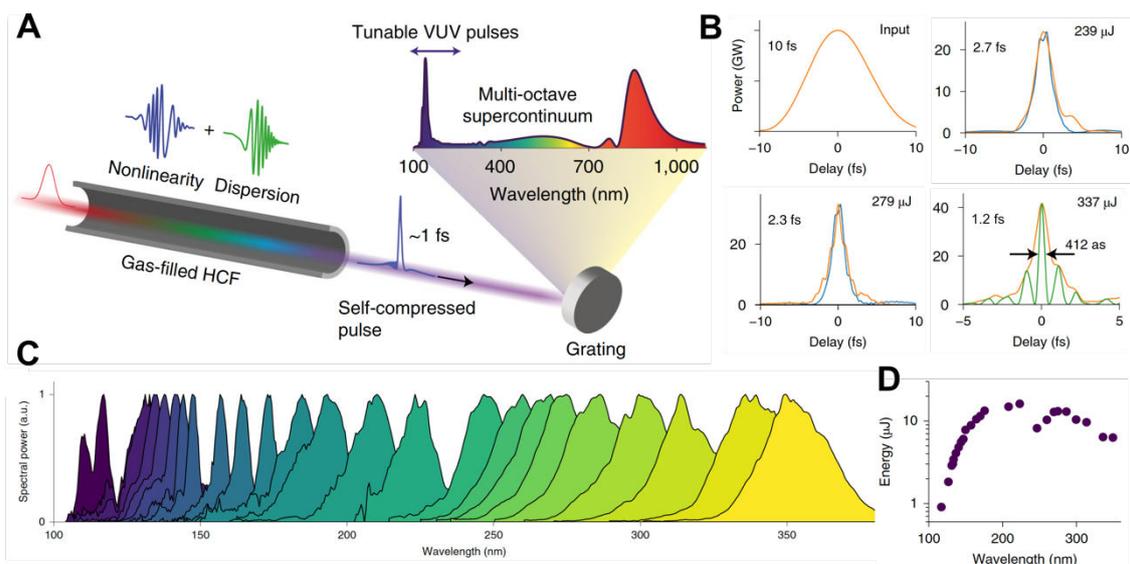

**Fig. 5.2** Generation of self-compressed, tuneable UV-to-VUV pulses in Helium-filled HCF. (A) Schematic of soliton self-compression using a Helium gas-filled HCF that is pumped by a 10 fs 800 nm pulse with pulse energy of 200-400 µJ. A self-compressed optical pulse is generated as a result of the soliton dynamics combining nonlinearity and dispersion. The spectrum of the self-compressed pulse is measured using a grating. (B) Self-compression of the 10-fs input pulse to between 2.7 fs and 1.2 fs is achieved using input pulse energies of 239 – 337 µJ. (C) Measured spectra containing differing central wavelengths between (122 - 350 nm) obtained with varied helium input gas pressure (0.23 – 4.0 bar). (D) The pulse energy of the self-compressed pulse as a function of its central wavelength. Figure adapted from [Travers2019].

Gas-filled single-ring PCFs have been shown to generate deep ultraviolet (DUV) few-femtosecond pulses with ~1 µJ pulse energy at MHz repetition rate, requiring only 20 µJ, <25 fs pump pulses [Köttig2017]. This is particularly useful for high-repetition rate UED/LIED (*i.e.* >100 kHz) which will require high-average power optical sources. DUV pulses through single-ring PCFs is generated through the emission of dispersive waves (DWs) from self-compressed solitons. Figs. 5.3A and 5.3B show the simulated temporal and spectral profiles of the self-compressed deep ultraviolet pulse, respectively, with 53 bar Helium-filled single-ring PCFs using 17 µJ pulse energy. Here within the fibre, the generated DW at ~200 nm was self-compressed down to <10 fs (FWHM) possessing a pulse energy of 2 µJ. Figs. 5.3C and 5.3D show the measured and simulated output UV pulse energy, respectively, for four different types of rare gases that fill the fiber. Here, the UV pulse energy is clearly dependent on both the identity of the rare gas and the input energy, with the best performance found to be with 53 bar of helium and an input pulse energy of >15 µJ [Köttig2017]. Utilizing such few-to-sub-fs tuneable UV-to-VUV pulses with 1 – 10 µJ pulse energies coupled with MIR-LIED has the potential to record attosecond "molecular movies" of, for example, proton transfer and roaming reactions which occur on the sub-100-fs timescale.

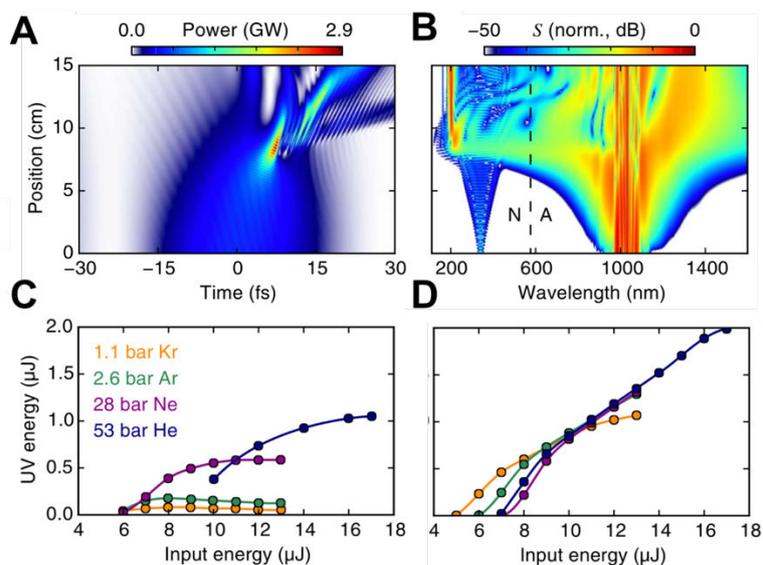

**Fig. 5.3** Gas-filled single-core PCF generation of deep ultraviolet few-femtosecond pulses with µJ pulse energy at MHz repetition rate. Simulated (A) temporal and (B) spectral profiles of the self-compressed deep ultraviolet pulses. The UV pulse energy (C) measured at 100 kHz and (D) simulated for UV pulses in the 200 – 220 nm range. Figure adapted from [Köttig2017].

Thirdly, as discussed in Section IVc, electron diffraction techniques that rely on the strong-field tunnel ionization of the sample to launch the recolliding electron must take into account that the ionization potential decreases with larger, more complex molecules. Consequently, a suitable balance must be struck between ionization saturation and ionizing the sample molecule to initiate the elastic rescattering process.

In the case of UED, improving the spatial resolution and the total temporal resolution can be achieved by using pulsed guns that generate electrons with relativistic energies at MHz repetition rates, which is a significant improvement on the currently used 120 Hz relativistic electron source [Weathersby2015, Filippetto2016, Schippers2019]. Employing a MHz relativistic electron gun in time-resolved MeV UED measurements will subsequently require a MHz high-average power optical source which is used to photoexcite the target sample. Such a MHz optical source will require an average power of greater than a kW [Schippers2019].

For both LIED and UED, more advanced structural retrieval algorithms are needed to successfully retrieve the geometric structure of more complex biomolecules, such as azo complexes or peptides. This is particularly important for time-resolved measurements which will possess multiple transient structures that may have non-negligible contributions to time-resolved electron diffraction data. Current retrieval algorithms (*e.g.* QRS and IAM theories) may not be tractable for such large molecular systems, requiring the extrema to be found in a multi-dimensional solution space. One possible alternative in the case of LIED, as discussed earlier, would be the use of FT-LIED, which does not rely on *a priori* knowledge of molecular structure or retrieval algorithms.

Finally, user-facilities based on field-free and field-dressed electron diffraction techniques similar to X-ray free-electron laser (XFEL) user-facilities that are currently offered will lead to a greater amount of interdisciplinary research between physics, chemistry and biology. The SLAC MeV UED user facility now offers external user experimental runs through a proposal selection procedure similar to that of XFELs, enabling MeV UED research in material and chemical systems [SLAC2019]. Both user-based facilities (*e.g.* XFELs, SLAC MeV UED) are highly complementary to small-scale laboratory-based techniques (*e.g.* LIED, keV UED, HHG), with both cases pushing one another to achieve the ultimate space and time resolution in time-resolved studies of biologically-relevant, complexes gas-phase molecules.

**Acknowledgements.** K.A. and J.B. acknowledge financial support from the Spanish Ministry of Economy and Competitiveness (MINECO), through the "Severo Ochoa" Programme for Centres of Excellence in R&D (SEV-2015-0522) Fundació Cellex Barcelona and the Centres de Recerca de Catalunya (CERCA) Programme/Generalitat de Catalunya, the European Research Council (ERC) for ERC Advanced Grant TRANSFORMER (788218), MINECO for Plan Nacional FIS2017-89536-P, Agència de Gestió d'Ajuts Universitaris i de Recerca for 2017 SGR1639, and Laserlab-Europe (EU-H2020 654148). We also acknowledge the Polish

National Science Center within the project Symfonia, 2016/20/W/ST4/00314, and the Marie Sklodowska-Curie Grant Agreement 641272.